%% file: paper_jrnl_clean.tex
\documentclass[journal]{IEEEtran}
\usepackage{amsmath,graphicx,algorithm,algorithmic}

\usepackage{amsmath,amssymb,amsthm,amsfonts,latexsym,bbm,xspace,graphicx,float,mathtools,paralist,verbatim,xcolor} 
\usepackage{url}
\PassOptionsToPackage{hyphens}{url}\usepackage{hyperref}

\input{macrodefs}

\newtheorem{definition}{Definition}
\newtheorem{theorem}{Theorem}

\newtheorem{assumption}{Assumption}

\title{
How can a Radar Mask its Cognition? 
\thanks{Short versions containing  partial results  appear in the IEEE International Conference on Acoustics, Speech and Signal Processing (ICASSP), 2022, International Conference of Information Fusion (FUSION), 2022 and IEEE International Conference on Decision and Control (CDC), 2022. }}
\author{Kunal Pattanayak,~\IEEEmembership{Student Member,~IEEE}, Vikram Krishnamurthy,~\IEEEmembership{Fellow,~IEEE} 
and Christopher Berry
\thanks{V. Krishnamurthy and K. Pattanayak are with the School of Electrical and Computer Engineering, Cornell University, Ithaca, New York, 14853 USA. e-mail: vikramk@cornell.edu, kp487@cornell.edu. C. Berry is with Lockheed Martin Advanced Technology Laboratories, Cherry Hill, NJ, 08002 USA. e-mail: christopher.m.berry@lmco.com. 
  This research was supported in part by a research contract from  Lockheed Martin,   the Army Research Office grant W911NF-21-1-0093 and the Air Force Office of Scientific Research grant  FA9550-22-1-0016.}}

\begin{document}

\maketitle

\begin{abstract}
A cognitive radar is a constrained utility maximizer that adapts its sensing mode in response to a changing environment. If an adversary can estimate the utility function of a cognitive radar, it can determine the radar's sensing strategy and mitigate the radar performance via electronic countermeasures (ECM). This paper discusses how a cognitive radar can {\em hide} its strategy from an adversary that detects cognition. The radar does so 
by transmitting purposefully designed sub-optimal responses
to spoof the adversary's Neyman-Pearson detector. We provide theoretical guarantees by ensuring the Type-I error probability of the adversary's detector exceeds a pre-defined level for a specified tolerance on the radar's performance loss. We illustrate our cognition masking scheme via  numerical examples involving waveform adaptation and beam allocation. We show that small purposeful deviations from the optimal strategy of the radar confuse the adversary by significant amounts, thereby masking the radar's cognition.  Our approach uses novel ideas from revealed preference in microeconomics and adversarial inverse reinforcement learning. Our proposed algorithms  provide a principled approach for system-level electronic counter-countermeasures (ECCM) to mask the radar's cognition, i.e.\,, hide the radar's strategy from an adversary.
We also provide performance bounds for our cognition masking scheme when the adversary has misspecified measurements of the radar's response.

\end{abstract}
\allowdisplaybreaks

\section{Introduction}

In abstract terms, a cognitive radar is a  constrained utility maximizer with multiple sets of utility functions and constraints that allow the radar to deploy different strategies depending on changing environments. Cognitive radars adapt their waveform scheduling and beam allocation by optimizing their utility functions in different situations. If a smart adversary can estimate the utility function or constraints of the cognitive radar, then it can exploit this information to mitigate the radar's performance (e.g., jam the radar with purposefully designed interference). A natural question is: {\em how can a cognitive radar hide its cognition from an adversary?} Put simply, how can a smart sensor hide its strategy  by acting dumb? We term this cognition-masking functionality as meta-cognition.\footnote{``Meta-cognition'' \cite{MI20} is used to describe a sensing platform  that switches between multiple objectives (constrained  utility functions).} A meta-cognitive radar~\cite{MI20} switches between multiple objectives (plans) to maintain stealth; for example, it can switch between the conflicting objectives of maximizing the signal-to-noise ratio of a target to maximizing privacy of its plan to maintain stealth.

A meta-cognitive radar pays a penalty for stealth - it deliberately transmits sub-optimal responses to keep its strategy hidden from the adversary resulting in performance degradation. This paper investigates 
how  a cognitive radar hide its strategy when the adversary observes the radar's responses. Our meta-cognition results are inspired by privacy-preserving mechanisms in differential privacy and adversarial obfuscation in deep learning with related works discussed below. Although this paper is radar-centric, we emphasize that the problem formulation and algorithms  also  apply to {\em adversarial inverse reinforcement learning} in general machine learning applications, namely, how to purposefully choose suboptimal actions to hide a strategy.

\subsection*{Related Works}
Cognitive radars are widely studied~\cite{cogradar_1,cogradar_2}. More recently, our papers~\cite{KAEM20,KPGKR21} deal with inverse reinforcement learning (IRL) algorithms for cognitive radars, namely, how can an adversary estimate the utility function of a cognitive radar by observing its decisions. 
Reconstructing a decision maker's utility function by observing its actions is the main focus of  IRL~\cite{NG00,ABB04,ZB08} in machine learning and revealed preference~\cite{Afr67,CD15} in micro-economics literature. In the radar literature, such IRL based adversarial actions to mitigate the radar's operations are called electronic countermeasures (ECM)~\cite{BD78,Kup17,KAEM20}. This paper builds on \cite{KAEM20, KPGKR21} and develops electronic counter-countermeasures (ECCM)~\cite{ECCM1,ECCM2,ECCM3}
to mitigate ECM. This paper assumes that adversary's ECM is unaware if the radar has ECCM capability, which is consistent with state-of-the-art ECCM literature.
The central theme of this paper is to apply results from revealed preference in micro-economics theory~\cite{Afr67,Var12}. To the best of our knowledge, this approach for ECCM to  hide  cognition  is novel.

Several works in literature~\cite{KH12,AM15,RO16} highlight how an adversary benefits from learning the radar's utility function. In \cite{KH12}, the adversary optimize its probes to increase the power of its statistical hypothesis test for utility maximization. \cite{AM15,RO16} show how revealed preference-based IRL techniques can be used to manipulate consumer behavior. 

In the radar context, \cite{iirl_naive} uses the Laplacian mechanism for meta-cognition; the cognitive radar anonymizes its trajectories via additive Laplacian noise. In our cognition masking approach, the radar mitigates adversarial IRL via purposeful perturbations from optimal strategy, where the perturbations are computed via stochastic gradient algorithms (see Algorithm~\ref{alg:noise_irp_utility} in Sec.\,\ref{sec:noisy_iirl}).

\subsection*{Context}

{\em Radar Design Paradigm: Cognition Masking vs LPI}. Low-probability-of-intercept (LPI) radars~\cite{LPI1,LPI2,LPI3} achieve stealth by minimizing the probability of the radar signals being detected by an adversarial target. Our rationale for stealth in this paper is at a higher level of abstraction than classical LPI.  The cognitive radar's aim is to {\em confuse} the adversary's detector, i.e.\,, ensure the adversary incorrectly reconstructs the radar's strategy with high probability.

{\em System level ECCM vs Pulse level ECCM.} Our cognition masking algorithm is implemented at the system level (Bayesian tracker level) and not the pulse level (Wiener filter level).
Pulse-level ECCM~\cite{PULSE1,PULSE2,PULSE3} accomplishes LPI-type functionalities for cognitive radars. Cognition masking hides the radar's {\em strategy} from the adversary instead of mitigating the adversary's detection of the radar's transmission. Hence, cognition masking ECCM is deployed at a higher level of abstraction than pulse level ECCM.

{\em Hiding Cognition against Optimal IRL vs Sub-optimal IRL.} The cognition masking results in this paper assume the adversary performs optimal IRL using Afriat's theorem~\cite{Afr67,Var12}. Afriat's theorem achieves optimal IRL for non-parametric utility estimation of a cognitive radar as it generates a polytope of {\em all viable utilities} that rationalizes a finite dataset of adversarial probes and radar responses.
However, our cognition masking results can be extended to {\em any} potentially sub-optimal IRL algorithm that generates a set-valued estimate of the radar's utility, as long as the radar has knowledge of the IRL algorithm being used by the adversary.
Algorithm~\ref{alg:arb-irl} in the appendix outlines how a cognitive radar can mask its cognition for an arbitrary IRL algorithm. At an abstract  level, cognition masking simply obfuscates a set-valued mapping from the adversary's dataset to a set of feasible utilities by intelligently distorting the radar's responses and hence, is not affected by the optimality of adversarial IRL. 

At a deeper level, this paper quantifies cognition masking performance when the adversary has misspecified measurements of the radar's response, and performs sub-optimal IRL. Theorem~\ref{thrm:misspec} (in  appendix) provides performance guarantees for cognition masking  when the radar does not know the misspecification errors and provides a bound on the cognitive masking performance in terms of the error magnitude.  \vspace{-0.3cm}\\

\subsection*{Why not an MDP or non-cooperative game?}
In machine learning based IRL~\cite{NG00,ZB08,RAT06}, the aim is to reconstruct the rewards of a Markov decision process  (MDP) subject to entropic constraints  on the policy. This requires complete   knowledge of the transition dynamics of the adversary's probes. In comparison, our radar-adversary interaction is batch-wise - the adversary transmits a batch of probe signals, and then the radar responds with a batch of responses.
This non-parametric identification of the radar's strategy is agnostic to transition dynamics in the adversary's probes. Hence, a static utility maximization setup is more realistic  for  IRL and inverse IRL involving cognitive radar.

We consider a radar-adversary interaction where the adversary is not aware of the radar's cognition masking strategy. 
A more general formulation is a Stackelberg game between the radar and the adversary, with the adversary as the leader and the radar as the follower. However, such an approach for computing the optimal meta-cognition strategy for the radar is ill-posed since  the existence of a pure and unique Nash equilibrium is not guaranteed. Finally, from an inverse game theoretic perspective, identifying if the radar-adversary behavior is consistent with Nash equilibrium is  intractable since the analyst needs to know both the radar's and adversary's utility function.
Addressing these issues is beyond the scope of this paper, and the subject of future work.
\vspace{-0.2cm}

\subsection*{Outline and Organization of Results}
\noindent {\em (i) Background. Inverse reinforcement learning (IRL):} In Sec.\,\ref{sec:background}, we formulate the interaction between a cognitive radar and an adversary target. We review the main idea of  revealed preference-based adversarial IRL algorithms, namely, Theorems~\ref{thrm:rp} and \ref{thrm:rp_constraint} used by the adversarial target to reconstruct the radar's strategy from its actions. Then we outline two examples, namely waveform adaptation and beam allocation.
Theorem~\ref{thrm:rp_vec_g} stated in Appendix~\ref{appdx:vector-valued-g} extends adversarial IRL to the case where the cognitive radar faces multiple constraints. Theorem~\ref{thrm:rp_vec_g} is omitted from the main text for readability.

\noindent {\em (ii) Masking Radar's Strategy from Adversarial IRL:} Sec.\,\ref{sec:iirl} contains our main meta-cognition results, namely, Theorems~\ref{thrm:irp} for mitigating adversarial IRL by masking the radar's strategy. 
The key idea is for the radar to deliberately deviate from its optimal (naive) response to ensure:\\ (1) its true strategy almost fails to rationalize its perturbed responses (masked from adversarial IRL), and\\
(2) its performance degradation due to sub-optimal responses does not exceed a particular threshold.
Theorem~\ref{thrm:irp_vec_g}  in Appendix~\ref{appdx:vector-valued-g} extends Theorem~\ref{thrm:irp} to the case where the cognitive radar has multiple constraints. Theorem~\ref{thrm:misspec} provides performance bounds on the cognition masking scheme of Theorem~\ref{thrm:irp} when the adversary has misspecified measurements of the radar's response.
\vspace{0.1cm}\\
{\em (iii) Masking Radar's Strategy from Adversarial IRL Detectors in noise.} 
Sec.\,\ref{sec:stoch_irl} extends our IRL and cognition masking results to the case where the adversary has noisy measurements of the radar's response. First, we define IRL detectors (Definition~\ref{def:noise_rp}) that {\em detect} radar's cognition in noise. Then, we enhance our cognition masking scheme of Theorem~\ref{thrm:irp} to mitigate the IRL detectors. The radar's cognition masking objective now is to 
maximize the detectors' conditional Type-I error probability, subject to a bound on its deliberate performance degradation.
\vspace{0.1cm}\\
{\em (iv) Numerical illustration of masking cognition by meta-cognitive radars.} 
Sec.\,\ref{sec:numerical_results} illustrates our meta-cognition results on two  target tracking functionalities, namely,  
waveform adaptation and beam allocation. 
Our numerical experiments show that the meta-cognition algorithms in this paper can effectively mask both the radar's utility function and resource constraint when the cognitive radar is probed by the adversarial target. Our main finding is that a small deliberate performance loss of the meta-cognitive radar suffices to mask the radar's strategy from the adversary to a large extent.

\section{Background. IRL to Estimate  Cognitive Radar} \label{sec:background}

Since this  paper investigates how to construct a cognitive radar that hides its utility from an adversarial IRL system, this section gives the background on how an adversarial system can use IRL to estimate the radar's utility.
An important aspect of the IRL framework below is that it is a necessary and sufficient condition for identifying cognition (utility maximization behavior); hence it can be considered as an  optimal IRL scheme.
Appendix~\ref{appdx:arb-irl} and \ref{appdx:misspec}  discuss  cognition masking when the adversary performs sub-optimal IRL. 

\subsection{Radar-Adversary Dynamics}
\begin{definition}[Radar-Target Interaction]\label{def:radar-adversary-interaction}  The cognitive radar-adversary interaction has the following dynamics:
\begin{equation}
\begin{split}
\text{target probe: } \probe_\time & \in\reals_+^d\\
\text{radar action: }\response_\time &  \in\reals_+^d\\
  \text{target state: }    \state_k & = \{\state_\time(t),~t=1,2,\ldots\}, \\
 \state_\time(t+1) & \sim  \pdf_{\probe_\time}(\state | \state_{k}(t)), ~\state_0 \sim \belief_0 \\
\text{radar observation: }     \obs_k  &\sim  \pdf_{\response_\time}(\obs | \state_\time)\\
\text{radar tracker: }    \belief_k &= \filter(\belief_{k-1}, \obs_k)\\
\text{observed radar action: } \nresponse_\time &= \response_\time + \resnoise_\time,~\resnoise_\time\sim f_{\resnoise}
  \end{split} \label{eq:model}
\end{equation}
\end{definition}
\noindent {\em Remarks.} We now give examples for the abstract model~\eqref{eq:model}.\\
\noindent 1. A widely used example~\cite{BLK08,LJ03} for the radar-adversary dynamics model~\eqref{eq:model} is that of linear Gaussian dynamics for target kinematics and linear Gaussian measurements:
\begin{align}
x_\time(t+1) &= A x_\time(t) + w_t(\probe_\time), \quad x_\time(0) \sim \pi_0 = \mathcal{N}(\hat{x}_0, \Sigma_0)
\nonumber\\
y_\time(t) &= C x_\time(t) + v_t(\response_\time),\quad\time=1,2,\ldots,\horizon\label{eqn:kalman-sys}
\end{align}
Here $x_\time(t) \in \mathcal{X} = \mathbb{R}^X$, $\obs_\time(t)\in\mathcal{Y} = \reals^Y$. $A$ is a block diagonal matrix \cite{BP99} when the target state represents its position and velocity in Euclidean space. The variables $w_t \sim \mathcal{N}(0, Q(\probe_\time))$ and $v_t \sim \mathcal{N}(0, R(\response_\time))$ are mutually independent Gaussian noise processes.
\\
\noindent 2. In this paper, we are only concerned with the asymptotic statistics of the radar tracker $\filter$~\eqref{eq:model} for our cognition-masking algorithms. One example is that of a Bayesian tracker (Kalman filter) where the asymptotic covariance of the state estimate is the unique positive semi-definite solution of the algebraic Riccati equation (ARE). Other tracker examples include the particle filter, interacting multiple model (IMM) filter etc.

We now proceed to define a cognitive radar which we assume in this paper to be a constrained utility maximizer.
\begin{definition}[Cognitive Radar] Consider the radar-adversary interaction dynamics of Definition~\ref{def:radar-adversary-interaction}. The cognitive radar chooses its response $\idresponse_\time$~\eqref{eq:model} at time $\time$ 
by maximizing a utility function $\utilityrad(\probe_\time,\cdot)$ subject to constraint $\nonlinbrad(\probe_\time,\cdot)\leq 0$:
    \begin{equation}\label{eqn:abstract_cog_radar}
    \begin{split}
    \idresponse_\time  \in \argmax~& \utilityrad(\probe_\time,\response),\\
    \nonlinbrad(\probe_\time,\response)&\leq 0,
    \end{split}
\end{equation}
We assume that $\nonlinbrad(\cdot)$ is an increasing function of  $\response$.
\label{def:cognitive-radar}
\end{definition}
\noindent {\em Remarks.}\\
1. In the main text of this paper, we consider a single constraint. This is consistent with most works in cognitive radar literature which also assume a single operating constraint. For example, in \cite{Hayk06}, the cognitive radar is constrained by a bound on the target dwell time (monotone in the time the radar spends tracking each target). In \cite{Bell15}, the radar's constraint is a bound on the receiver sensor processing cost (monotone in the radar's choice of sensor accuracy for target tracking). Hence, we only consider the operating cost of the radar in the main text which is reflected in the radar's scalar-valued constraint $\nonlinbrad$ in \eqref{eqn:abstract_cog_radar}. \\
2. {\em Multiple resource constraints.} Our IRL methodology discussed below can be extended to multiple resource constraints ($\nonlinbrad$ is vector-valued). However, for readability, we only consider a scalar-valued constraint $\nonlinbrad$ in the main text of this paper. We consider multiple resource constraints in Appendix~\ref{appdx:vector-valued-g}. The notation for IRL and cognition masking results is complicated for vector-valued $\nonlinbrad$ and hence omitted from the main text.

\subsection{Adversarial IRL for Identifying Strategy of Cognitive Radar}
We now review the main results for adversarial IRL, namely, how an adversary can identify and reconstruct the radar's strategy by observing the radar's responses. The adversarial IRL system is schematically shown in Fig.\,\ref{fig:schematic-adversarial-irl}. The key idea is to formulate the adversary's task of identifying the radar's strategy as a linear feasibility problem in terms of the radar's responses. This paper considers two distinct scenarios in terms of the dependency of the adversary's probe $\probe_\time$ on the radar's utility $\utilityrad$ and resource constraint $\nonlinbrad$ in \eqref{eqn:abstract_cog_radar}. The two scenarios are formalized in Assumptions~\ref{asmp:utility} and \ref{asmp:constraint} below in our IRL results,  
Theorems~\ref{thrm:rp} and \ref{thrm:rp_constraint},  and justified in Sec.\,\ref{sec:radar_tracking_examples} in the tracking examples of waveform adaptation and beam allocation.

\subsection*{IRL for Identifying Utility Function}
Theorem~\ref{thrm:rp} below provides a set-valued reconstruction algorithm to estimate the radar's utility function when the adversary controls the radar's resource constraint. Such scenarios where the adversary knows the radar's resource constraint is formalized below in Assumption \ref{asmp:utility}:
\begin{assumption}
    The radar's resource constraint $\nonlinbrad(\cdot)$ in \eqref{eqn:abstract_cog_radar} is linear in the adversary's probe $\probe_\time$ and the radar's utility $\utilityrad(\cdot)$ is independent of $\probe_\time$:
    \begin{equation}\label{eqn:special_case_cog_radar_utility}
    \nonlinbrad(\probe_\time,\response) = \probe_\time'\response-1,~\utilityrad(\probe_\time,\response) \equiv \utilityrad(\response)
    \end{equation}
    \noindent \underline{IRL objective.} The adversary aims to reconstruct the radar's utility $\utilityrad(\cdot)$ using the dataset $\dataset_\nonlinb$, where $\dataset_\nonlinb$ is defined as:
    \begin{equation}\label{eqn:dataset_IRL_utility}
        \dataset_\nonlinb = \{\nonlinbrad(\probe_\time,\cdot),\response_\time\}_{\time=1}^\horizon,
    \end{equation}
    where $\nonlinbrad(\probe_\time,\cdot)$ is defined in \eqref{eqn:special_case_cog_radar_utility}.    
\label{asmp:utility}
\end{assumption}

Let us now state Theorem~\ref{thrm:rp} for achieving IRL when assumption~\ref{asmp:utility} holds.
\begin{theorem}[IRL for Identifying Radar's Utility Function]\label{thrm:rp} Consider the cognitive radar described in Definition~\ref{def:radar-adversary-interaction}. Suppose assumption \ref{asmp:utility} holds. Then:\\
(a) The adversary checks for the existence of a feasible utility function that satisfies \eqref{eqn:abstract_cog_radar} by checking the feasibility of a set of linear inequalities:
\begin{equation}
\begin{split}
   & \text{There exists a feasible }\param\in\reals_+^{2\horizon}~\text{s.t. }\AFT(\param,\dataset_\nonlinb)\leq \mathbf{0},\\
  \Leftrightarrow  &\exists~\utility~\text{s.t. } \response_\time \in \argmax \utility(\response),~\probe_\time'\response\leq 1~\forall\time,
\end{split}\label{eqn:abstract_IRL_utility}
\end{equation}
where dataset $\dataset_\nonlinb$ is defined in \eqref{eqn:dataset_IRL_utility} and the set of inequalities $\AFT(\cdot)\leq 0$ is defined in Appendix~\ref{appdx:IRL_inequalities}.\\
\noindent (b) If $\AFT(\cdot,\dataset_\nonlinb)$ has a feasible solution, the set-valued IRL estimate of the radar's utility $\utilityrad$ is given by:
\begin{equation}
\begin{split}
\utility_{\IRL}(\response) & \equiv \{\utility_{\IRL}(\response;\param) : \AFT(\param,\dataset_\nonlinb)\leq \mathbf{0}\},\\
    \utility_{\IRL}(\response;\param) & = \underset{\time\in \{1,2,\dots,\horizon\}}{\operatorname{min}}\{\param_\time+\param_{\time + \horizon}~\probe_\time'(\response-\response_\time)\}.
    \end{split}
    \label{eqn:estutility}  
\end{equation}
\end{theorem}

Theorem~\ref{thrm:rp} is well known in micro-economics as Afriat's theorem~\cite{Afr67,Var12} and widely used for set-valued estimation of consumer utilities from logged offline data. In complete analogy, the adversary also performs IRL on a batch of probe-response exchanges with the cognitive radar to reconstruct the radar's utility
\footnote{Afriat's theorem with linear constraints \eqref{eqn:special_case_cog_radar_utility} has been generalized to non-linear monotone constraints in literature~\cite{FM09}. For the radar context in this paper, it suffices to assume a linear constraint when the adversary is trying to estimate the radar's utility}.
Abstractly, Theorem~\ref{thrm:rp} says that given a finite dataset, the adversary can at best construct a polytope of feasible strategies that rationalize the adversary's dataset. Theorem~\ref{thrm:rp} achieves IRL when the radar faces a single operating constraint. We discuss adversarial IRL for multiple resource constraints in Theorem~\ref{thrm:rp_vec_g} in Appendix~\ref{appdx:vector-valued-g}. Then the linear feasibility test of \eqref{eqn:abstract_IRL_utility} generalizes to a mixed-integer linear feasibility test, linear in the real-valued feasible variables in the multi-constraint case.

\subsection*{IRL for Identifying Radar's Resource Constraints} 
In certain scenarios, the utility of the radar is well known (e.g., signal-to-noise ratio), but the operational constraints of the radar are not known. We formalize such scenarios where the adversary knows the radar's utility function below as Assumption~\ref{asmp:constraint}:
\begin{assumption}\label{asmp:constraint}
The radar's utility function $\utilityrad(\cdot)$~\eqref{eqn:abstract_cog_radar} is controlled by the adversary's probe $\probe_\time$, the radar's resource constraint $\nonlinbrad$ is independent of $\probe_\time$ and has the following form:
    \begin{equation}\label{eqn:special_case_cog_radar_constraint}
        \nonlinbrad(\probe_\time,\response) \equiv \nonlinbrad(\response)-\thresh_\time,~\thresh_\time>0,
    \end{equation}
    where $\thresh_\time,\nonlinb$ are independent of $\probe_\time$.\\
    \underline{IRL objective.} The adversary aims to reconstruct $\nonlinbrad(\cdot)$ using the dataset $\dataset_\utility$, where $\dataset_\utility$ is defined as:
    \begin{equation}\label{eqn:dataset_IRL_constraint}
        \dataset_u = \{\utilityrad(\probe_\time,\cdot),\response_\time\}_{\time=1}^\horizon.
    \end{equation}
\end{assumption}
IRL for estimating the radar resource constraints
has the same structure as that of Theorem~\ref{thrm:rp} and is discussed in the appendix. IRL for Assumption~\ref{asmp:constraint} is formally stated in Theorem~\ref{thrm:rp_constraint} in Appendix~\ref{appdx:IRL_constraint} and summarized below:
\begin{align}
&    \nonlinb_{\IRL}(\response)  \equiv \{\nonlinb_{\IRL}(\response;\param):\AFT(\param,\dataset_\utility)\geq\mathbf{0}\},\label{eqn:summary_IRL_constraint}\\
    &\nonlinb_{\IRL}(\response;\param) =\underset{\time\in\{1,2,\ldots,\horizon\}}{\max}\{\param_\time + \param_{\horizon+\time}(\utilityrad(\probe_\time,\response)-\utilityrad(\probe_\time,\response_\time))\},\nonumber
\end{align}
where $\nonlinb_{\IRL}$ is the adversary's set-valued estimate of the radar's constraint $\nonlinbrad$, dataset $\dataset_\utility$ is defined in \eqref{eqn:dataset_IRL_constraint} and $\param\in\reals_+^{2\horizon}$ is a feasible vector wrt the feasibility test $\AFT(\cdot,\dataset_\utility)\geq \mathbf{0}$. Note how the IRL feasibility inequalities in \eqref{eqn:summary_IRL_constraint} are identical to that of \eqref{eqn:abstract_IRL_utility} in Theorem~\ref{thrm:rp} but with the inequality direction reversed.

\subsection{Examples of IRL for Identifying Radar Cognition}\label{sec:radar_tracking_examples}
Below, we discuss two examples of cognitive radar functionalities, namely, waveform adaptation and beam allocation. Throughout this paper, we will use the two examples below for contextualizing our cognition masking results.

\subsubsection{Example 1. Waveform Adaptation for Cognitive Radar}\label{sec:waveform}
Waveform adaptation~\cite{LI17,KW94,HK08,GL09} is a crucial functionality of a cognitive radar. Consider a cognitive radar with linear Gaussian dynamics and measurements~\eqref{eqn:kalman-sys}. The cognitive radar's aim is to choose the optimal sensor mode (observation noise covariance) based on the target's maneuvers. A more accurate sensor results in more precise observations, but is also costlier to deploy. Appendix~\ref{appdx:waveform} formalizes the optimal waveform adaptation and abstracts the problem as the constrained utility maximization problem of \eqref{eqn:abstract_cog_radar}. The key idea is to assume that adversary's probe $\probe_\time$  and radar's response $\response_\time$ are the eigenvalues of covariance matrices $\snoisecov$ and $\onoisecov^{-1}$, respectively, and hence, parameterize the state and observation noise covariance in the state space model of \eqref{eqn:kalman-sys}. Appendix~\ref{appdx:waveform} then shows the equivalence between an upper bound on the radar's asymptotic covariance $(\Sigma^\ast(\probe_\time,\response_\time))^{-1}$ and the linear constraint $\probe_\time'\response\leq 1$. In summary, the cognitive radar's optimal waveform adaptation strategy can be abstracted as:
\begin{equation}
    \response_\time \in \argmax~ \utilityrad(\response),~\probe_\time'\response\leq 1,\label{eqn:abstract_waveform}
\end{equation}
where $u$ is the radar's utility, and the linear constraint $\probe_\time'\response\leq 1$ equivalently bounds the {\em asymptotic precision} of the radar.

{\em IRL for optimal waveform adaptation.}
The adversary's aim is to identify the radar's utility function $u$. Also, the setup of 
~\eqref{eqn:abstract_waveform} falls under Assumption~\ref{asmp:utility}. Hence, the adversary uses the IRL test of \eqref{eqn:abstract_IRL_utility} in Theorem~\ref{thrm:rp} for identifying $\utility$.

\subsubsection{Example 2. Beam allocation for Cognitive Radar}\label{sec:beam}
Appendix~\ref{appdx:beam}  discusses  optimal beam allocation~\cite{KE01,Kri16,KD09,beam_1,beam_2,beam_3}.
The cognitive radar's aim is to allocate its beam intensity optimally between multiple targets. Compared to a target with less jerky maneuvers, a target with unpredictable maneuvers requires a more focused beam for the SNR to lie above a certain threshold. Appendix~\ref{appdx:beam} formalizes the beam allocation problem and abstracts the problem as a constrained utility maximization problem~\eqref{eqn:abstract_cog_radar}. The key idea is to relate the adversary's probe $\probe_\time$ to the asymptotic {\em predicted} precision of the radar tracker. In summary, the cognitive radar's optimal waveform adaptation problem can be abstracted as:
\begin{equation}
   \response_\time \in \argmax~\utilityrad(\probe_\time,\response) \equiv \prod_{i=1}^\probedim \response(i)^{\probe_\time(i)},~\|\response\|_\paramg\leq \thresh_\time,\label{eqn:abstract_beam}
\end{equation}
where the radar maximizes a Cobb-Douglas utility subject to a bound $\thresh_\time$ on the total transmit beam intensity ($\paramg$-norm of intensity vector) for all $\time$.

{\em IRL for optimal beam allocation.} Since the adversary knows the radar's utility (Assumption~\ref{asmp:constraint}), its aim is to identify the radar's constraint $\nonlinbrad(\cdot)-\thresh_\time\leq 0$ using the IRL test \eqref{eqn:abstract_IRL_constraint} in Theorem~\ref{thrm:rp_constraint}.

{\bf Summary}. This section discussed how an adversary can deploy IRL to estimate a cognitive radar's utility and constraint.
While IRL with a single operational constraint is discussed in~\cite{KAEM20}, the IRL algorithm for multiple constraints (in  Appendix~\ref{appdx:vector-valued-g}) is new.

\section{Inverse IRL. Masking Radar Utility and Constraints  from Adversarial IRL} \label{sec:iirl}

Having discussed how an IRL system can detect a cognitive radar, 
we are now ready to design a cognitive radar that is aware of the adversary's IRL motives and  hides its strategy (utility function and resource constraints) from the IRL system. In radar terminology, IRL for mitigating a radar system falls under the field of electronic counter measures (ECM). Since meta-cognition deals with spoofing adversarial IRL, it can be viewed as a form of electronic counter-counter measure (ECCM) against ECM.

\begin{figure}[ht]
    \centering
    \includegraphics[width = \columnwidth]{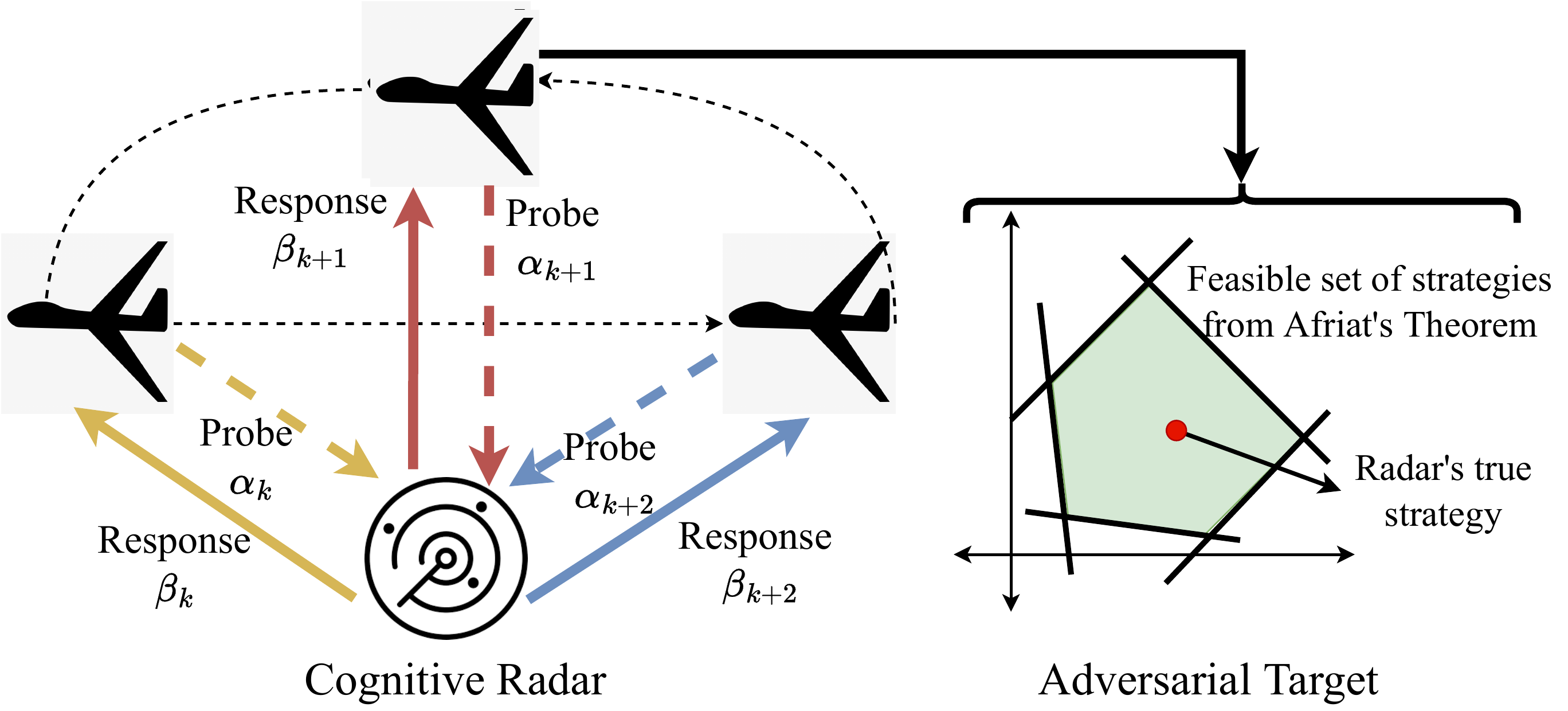}
    \caption{Schematic of adversarial IRL against cognitive radars.
    The adversary observes a sequence of decisions of the cognitive radar in response to a sequence of adversarial probes. Revealed preference-based adversarial IRL (Afriat's Theorem)~\cite{Afr67,Var12} is equivalent to checking the existence of a feasibility polytope for a set of inequalities (Afriat's Theorem~\cite{Afr67,Var12}).
    Our aim in this paper is to make adversarial IRL cumbersome - how to purposefully distort radar responses meta-cognition objective in this paper is to spoof adversarial IRL, namely, how to make checking linear feasibility difficult.}
    \label{fig:schematic-adversarial-irl}
\end{figure}

{\em Rationale. How to hide cognition?} 
Recall that the feasibility of \eqref{eqn:abstract_IRL_utility},~\eqref{eqn:abstract_IRL_constraint} is both necessary and sufficient for identifying utility maximization behavior~\eqref{eqn:abstract_cog_radar}; see \cite{Afr67,Var12} for the proof. Hence, a cognitive radar's true strategy lies within the polytope of feasible strategies computed by the adversary (Fig.\,\ref{fig:schematic-adversarial-irl}). The cognition masking rationale in this paper is to transmit purposefully designed  perturbed responses that  ensure the radar's true strategy lies close to the edge of the polytope of feasible strategies. The distance from the edge of the feasibility polytope is a measure of goodness-of-fit of the strategy to the radar's responses; see Definitions~\ref{def:margin_utility}, \ref{def:margin_constraint} below. 
In other words, the radar deliberately sacrifices performance to ensure its strategy poorly rationalizes its perturbed responses, hence hiding its strategy from adversarial IRL.

\subsection*{Main Result. How a radar can mask its  utility/constraints}  
Theorem~\ref{thrm:irp} below is our main result for cognition masking. Theorem~\ref{thrm:irp} uses the concept of feasibility margin - how far is a strategy from failing the IRL feasibility tests~\eqref{eqn:abstract_IRL_utility} or \eqref{eqn:abstract_IRL_constraint}. We define two margins - $\margin_\utility$ and $\margin_\nonlinb$ for the feasibility margins of feasible utilities $\utility$ and constraints $\nonlinb$, respectively.
\begin{definition}[Feasibility Margin for Reconstructed Utility~\eqref{eqn:abstract_IRL_utility}] \label{def:margin_utility} Consider the dataset $\dataset_\nonlinb$ defined in \eqref{eqn:dataset_IRL_utility}. The feasibility margin $\margin_u(\dataset_\nonlinb)$ defined below measures how far is the utility $\utility$ is from failing the IRL feasibility test~\eqref{eqn:abstract_IRL_utility}.
\begin{align}
    \margin_{u}(\dataset_\nonlinb) = & \min_{\eps\geq 0}~\eps,~\AFT(\utility,\dataset_\nonlinb)+\eps\mathbf{1}\geq\mathbf{0},
    \label{eqn:margin_rp}
\end{align}
where $\mathbf{1}$ is the column vector of all ones.
\end{definition}

\begin{definition}[Feasibility Margin for Reconstructed Constraints~\eqref{eqn:abstract_IRL_constraint}] \label{def:margin_constraint} Consider the dataset $\dataset_\utility$ defined in \eqref{eqn:dataset_IRL_constraint}. The feasibility margin $\margin_g(\dataset_\utility)$ defined below measures how far the constraint $\nonlinb$ is from failing the IRL feasibility test~\eqref{eqn:abstract_IRL_constraint}:
\begin{align}
    \margin_{g}(\dataset_\utility) = & \min_{\eps\geq0}\eps,~\AFT(\nonlinb,\dataset_\utility) - \eps\mathbf{1}\leq\mathbf{0}, 
    \label{eqn:margin_rp_constraint}
\end{align}
where $\mathbf{1}$ is the column vector of all ones.
\end{definition}

The margin \eqref{eqn:margin_rp},~\eqref{eqn:margin_rp_constraint} is a measure of goodness-of-fit for the IRL feasibility inequalities \eqref{eqn:abstract_IRL_utility}~and \eqref{eqn:abstract_IRL_constraint}, respectively, for any feasible strategy.\footnote{
Strictly speaking, the margin~\eqref{eqn:margin_rp} is the minimum perturbation so that $\AFT(\utility_{\AFT},\dataset_\utility)$ is infeasible, where $\utility_{\AFT}$ is the finite-dimensional projection of $\utility$ for the IRL feasibility test defined in \eqref{eqn:IRL_reverse_map} in Appendix~\ref{appdx:IRL_inequalities}. However, we abuse notation and express the feasibility test as $\AFT(\utility,\dataset_\utility)$ for the sake of simplicity of exposition. We abuse notation in a similar way for \eqref{eqn:margin_rp_constraint}} If $\utility$ is a feasible utility that rationalizes $\dataset_\nonlinb$~\eqref{eqn:dataset_IRL_utility}, we have $\AFT(\utility,\dataset_\nonlinb)\leq \mathbf{0}$ from \eqref{eqn:abstract_IRL_utility}. Hence, the margin for $\utility$ is the minimum {\em non-negative} perturbation so that the IRL test of \eqref{eqn:abstract_IRL_utility} fails, that is, $\AFT(\cdot,\dataset_\nonlinb)+\eps\mathbf{1}\geq \mathbf{0}$. 
Similarly, if $\nonlinb$ is a feasible resource constraint that rationalizes $\dataset_\utility$~\eqref{eqn:dataset_IRL_constraint}, we have $\AFT(\utility,\dataset_\nonlinb)\geq \mathbf{0}$ from \eqref{eqn:abstract_IRL_constraint}. Hence, the margin for $\utility$ is the minimum {\em non-positive} perturbation so that the IRL test of \eqref{eqn:abstract_IRL_utility} fails, that is, $\AFT(\cdot,\dataset_\nonlinb)-\eps\mathbf{1}\geq \mathbf{0}$. Equivalently, the margin measures how far a strategy lies from the edge of the polytope of feasible strategies. \footnote{There exist several robustness measures in literature \cite{VAR_book91,robustness_measure_1,robustness_measure_2,robustness_measure_3,robustness_measure_4} that check how well a {\em dataset} satisfies economic based rationality. Our cognition masking aim is more subtle - our aim is to ensure a particular {\em strategy} rationalizes a dataset poorly by minimizing its feasibility margin~\eqref{eqn:margin_rp}~\eqref{eqn:margin_rp_constraint}.}
The concept of margins arises in many prominent areas of machine learning, for example, in support vector machines (SVM)~\cite{SVM} and also IRL; see \cite{RAT06} for max-margin IRL. In the radar context, a strategy with a large feasible margin is a high-confidence point estimate of the radar's strategy and hence, at higher risk of getting exposed.

We are now ready to state our first cognition masking result, Theorem~\ref{thrm:irp}. Theorem~\ref{thrm:irp} ensures the radar's true strategy has a low feasibility margin wrt the IRL tests of Theorems~\ref{thrm:rp}, \ref{thrm:rp_constraint} by deliberately perturbing the radar's naive responses~\eqref{eqn:abstract_cog_radar}. In a sense, the radar optimally switches between maximizing its performance and maximizing the privacy of its plan.

\begin{theorem}[Masking Cognition from Adversarial IRL Feasibility Tests.]
Consider the cognitive radar~\eqref{eqn:abstract_cog_radar} from Definition~\ref{def:cognitive-radar}. Let $\{\response_\time^\ast\}_{\time=1}^\horizon$ denote the naive response sequence~\eqref{eqn:abstract_cog_radar} that maximizes the cognitive radar's utility. Then:  \\
(i)~\underline{Masking Utility Function from IRL.} 
Suppose Assumption~\ref{asmp:utility} holds. The response sequence $\{\pertresponse_{1:\horizon}^\ast\}$ defined below masks the radar's utility $\utilityrad$ from the adversary by ensuring $\utilityrad$ passes the IRL feasibility test~\eqref{eqn:abstract_IRL_utility} with a sufficiently low margin~\eqref{eqn:margin_rp} parametrized by $\eta\in[0,1]$:\vspace{-0.3cm}
\begin{align}
 \{\pertresponse_{1:\horizon}^\ast\}& = \underset{\{\response_\time\geq \mathbf{0},~ \probe_\time'\response_\time\leq 1 \}}{\argmin} \sum_{\time=1}^\horizon \utilityrad(\idresponse_\time) - \utilityrad(\response_\time), \label{eqn:irp}\\
&\quad\quad\quad\quad\margin_\utilityrad(\dataset_\nonlinb) \leq (1-\eta)~\margin_\utilityrad(\dataset_\nonlinb^\ast),\label{eqn:constraint_lowmargin}
\end{align}
where  dataset $\dataset_\nonlinb^\ast=\{\probe_\time'(\cdot)-1,\idresponse_\time\}_{\time=1}^\horizon$ is the adversary's dataset when the radar transmits naive responses $\{\idresponse_\time\}_{\time=1}^\horizon$, and $\dataset_\nonlinb$ is defined in \eqref{eqn:dataset_IRL_utility}.\\
\noindent~(ii) \underline{Masking Resource Constraint from IRL.} Suppose Assumption~\ref{asmp:constraint} holds.
The response sequence $\{\pertresponse_{1:\horizon}^\ast\}$ defined below masks the radar's resource constraint $\nonlinbrad$ from the adversary by ensuring $\nonlinbrad$ passes the IRL feasibility test~\eqref{eqn:abstract_IRL_constraint} with a sufficiently low margin~\eqref{eqn:margin_rp_constraint} parametrized by $\eta\in[0,1]$:\vspace{-0.3cm}
\begin{align}
\{\pertresponse_{1:\horizon}^\ast\}& =\underset{\{\response_\time\geq \mathbf{0},~ \nonlinb(\response_\time)\leq \thresh_\time \}}{\argmin} \sum_{\time=1}^\horizon \utilityrad(\idresponse_\time) - \utilityrad(\response_\time), \label{eqn:irp_constraint}\\
&\quad\quad\quad\quad \margin_\nonlinbrad(\dataset_\utility) \leq (1-\eta)\margin_\nonlinbrad(\dataset_\utility^\ast),\label{eqn:constraint_lowmargin_budget}
\end{align}
where dataset $\dataset_\utility^\ast=\{\utilityrad(\probe_\time,\cdot),\idresponse_\time\}_{\time=1}^\horizon$ is the adversary's dataset when the radar transmits naive responses $\{\idresponse_\time\}_{\time=1}^\horizon$, and $\dataset_\utility$ is defined in \eqref{eqn:dataset_IRL_constraint}.
\label{thrm:irp}
\end{theorem}

\begin{figure*}
    \centering    \includegraphics[width=0.95\linewidth]{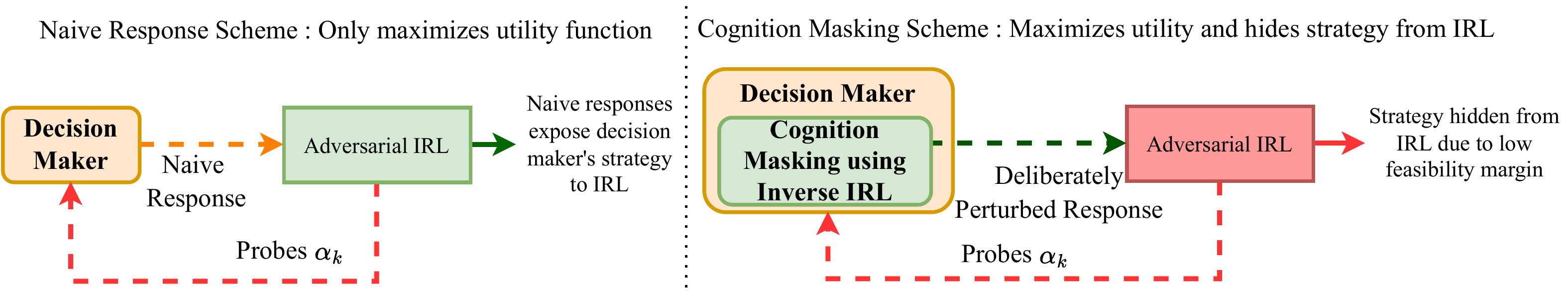} \caption{Schematic of the cognitive radar masking its strategy from adversarial IRL (via Theorem~\ref{thrm:irp}). \\
    {\em Naive response scheme (Left):} The adversary sends a sequence of probe signals to the radar and records its responses to the adversary's probes. The radar's strategy passes the IRL feasibility test of Theorem~\ref{thrm:rp} with a large margin if the radar transmits naive responses~\eqref{eqn:abstract_cog_radar} and can be reconstructed by IRL.\\
    {\em Cognition masking scheme (Right):} 
    If the radar is aware of adversarial IRL, the radar deliberately perturbs its responses according to Theorem~\ref{thrm:irp} to hide its strategy from the adversary at the cost of performance degradation. In Sec.\,\ref{sec:numerical_results}, we illustrate via numerical examples how small deliberate perturbations in the radar's naive responses mask the radar's strategy from adversarial IRL to a large extent.
    }
    \label{fig:schematic}
\end{figure*}

\begin{figure}
    \centering \includegraphics[width=0.6\columnwidth]{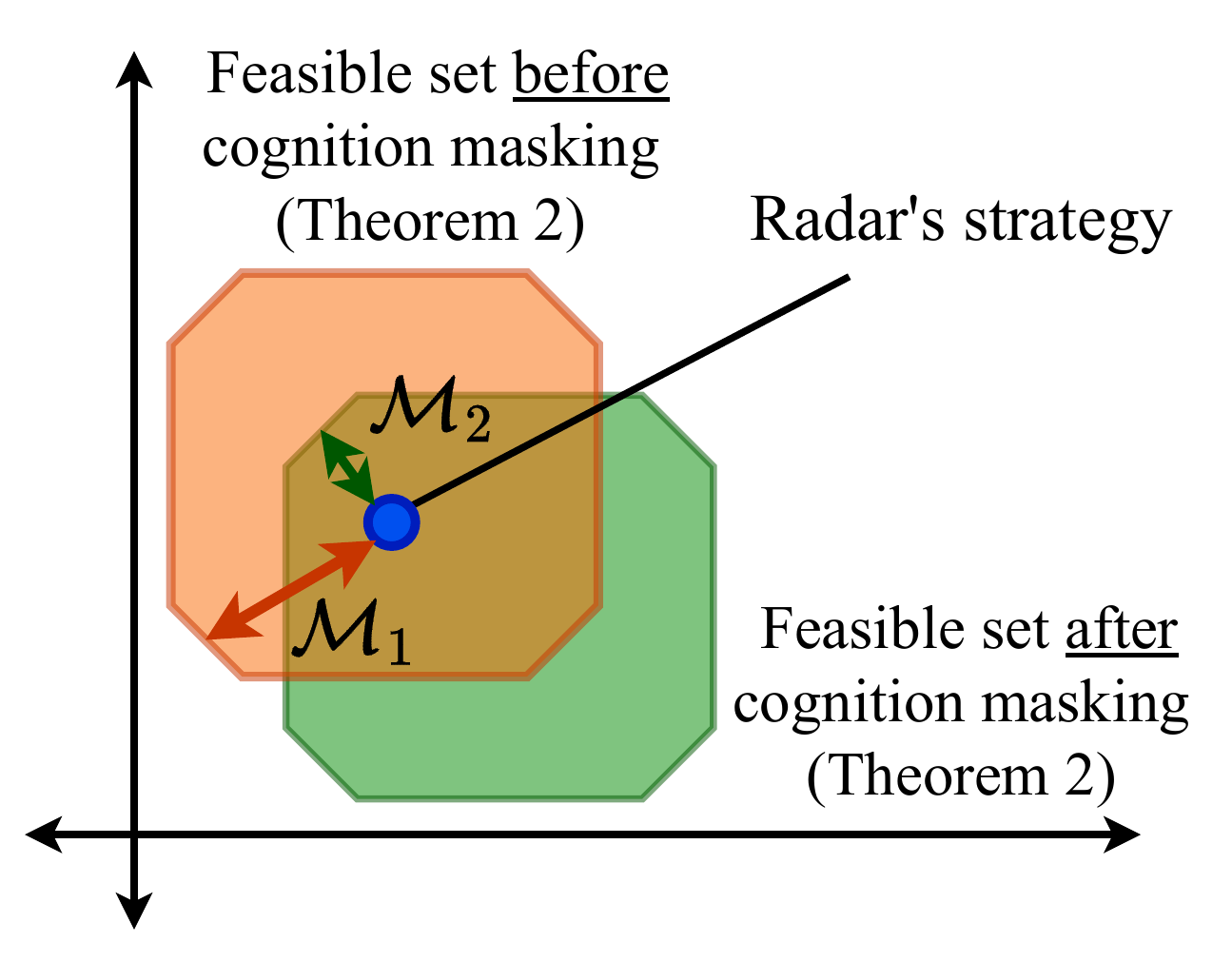}
    \caption{Cognition Masking for mitigating adversarial IRL. The radar's naive responses pass the IRL feasibility tests in Theorems~\ref{thrm:rp} and \ref{thrm:rp_constraint} with a large feasibility margin $\margin_1$. Cognition masking distorts the feasibility polytope so that the radar's true strategy is almost infeasible (low margin $\margin_2$) wrt the IRL feasibility inequalities (close to the edge of feasibility polytope). Hence, the true strategy is a low-confidence estimate for IRL and successfully hidden from the adversary.
    } 
    \label{fig:illth2}
\end{figure}

Theorem~\ref{thrm:irp} is our first result for masking cognition; see Algorithm~\ref{alg:irp} for a step-wise procedure for masking the radar's utility~\eqref{eqn:irp}. This is schematically illustrated in Fig.\,\ref{fig:illth2}. Theorem~\ref{thrm:irp} computes the {\em optimal} sub-optimal response of the radar that sufficiently mitigates adversarial IRL. The radar minimizes its performance degradation (maximizes {\em Quality-of-service (QoS)}) due to sub-optimal responses, subject to a bound~\eqref{eqn:constraint_lowmargin},~\eqref{eqn:constraint_lowmargin_budget} on the feasibility margin of the radar's strategy (maximizes {\em adversarial confusion}). Theorem~\ref{thrm:irp} can be viewed as an {\em inverse IRL} (I-IRL) scheme that mitigates an IRL system and is a critical feature of a {\em meta-cognitive} radar that switches between different plans. For completeness, Appendix~\ref{appdx:vector-valued-g} extends cognition masking to the case where the cognitive radar faces multiple constraints. Theorem~\ref{thrm:irp_vec_g} generalizes the cognition masking scheme of Theorem~\ref{thrm:irp} to the multi-constraint case where the adversary uses Theorem~\ref{thrm:rp_vec_g} for optimal IRL. Also, Appendix~\ref{appdx:misspec} discusses cognition masking when the adversary has {\em mis-specified} measurements of the radar's responses. Our key result is Theorem~\ref{thrm:misspec} that provides a performance bound on the cognition masking scheme of Theorem~\ref{thrm:irp} in terms of the misspecification error magnitude.

{\em Extent of cognition-masking $\eta$ in Theorem~\ref{thrm:irp}.}
A smaller value of $\eta$ implies a larger extent of cognition masking from adversarial IRL and also a greater degradation in the radar's performance. One extreme case is setting $\eta=0$. This results in maximal masking of the radar's strategy. That is, the IRL feasibility inequalities~\eqref{eqn:abstract_IRL_utility},~\eqref{eqn:abstract_IRL_constraint} are no more feasible and there exists {\em no feasible strategy} that rationalizes the radar's responses. Setting $\eta=0$ also causes the radar to deviate maximally from its naive responses~\eqref{eqn:abstract_cog_radar}, and hence results in a large performance degradation.
The other extreme case is setting $\eta=1$. In this case, the radar simply transmits its naive response~\eqref{eqn:abstract_cog_radar} and its strategy is not hidden from the adversary.

\begin{algorithm} \caption{Masking Radar's Utility via Theorem~\ref{thrm:irp} from IRL Feasibility Test~\eqref{eqn:abstract_IRL_utility}} \label{alg:irp}
Step 1. Compute radar's naive response sequence $\idresponse_{1:\horizon}$ by solving the convex optimization problem~\eqref{eqn:abstract_cog_radar}:
\begin{equation*}
    \idresponse_\time = \argmin \utilityrad(\response),~\nonlinbrad(\probe_\time,\response)\leq 0,~\response\geq\mathbf{0}~\forall\time\in\{1,2,\ldots,\horizon\},
\end{equation*}
where $\utilityrad$ is concave monotone in $\response$ and $\nonlinbrad(\probe_\time,\response)$ is convex monotone in $\response$.\\
Step 2. Choose $\eta\in[0,1]$ (extent of cognition masking from IRL feasibility test).\\
Step 3. Compute upper bound $\margin_{\operatorname{thresh}}$ on desired margin~\eqref{eqn:margin_rp} after cognition masking:\\ $\margin_{\operatorname{thresh}}=(1-\eta)~\margin_\utilityrad(\{\probe_\time,\idresponse_\time\}_{\time=1}^\horizon)$, where $\margin_{\utilityrad}$ is defined in \eqref{eqn:margin_rp}.\\
Step 4. Compute the cognition-making responses by solving the following optimization problem:
\begin{equation}\begin{split}s
\{\pertresponse_{1:\horizon}^\ast\}&_{\IIRLU}=\argmin~\sum_{\time=1}^\horizon \utilityrad(\idresponse_\time) - \utilityrad(\response_\time),\\
&\response_\time\geq \mathbf{0},~\probe_\time'\response_\time\leq  1~\forall\time\in\{1,2,\ldots,\horizon\},\\
&\margin_{\utilityrad}(\{\probe_\time,\response_\time\}_{\time=1}^\horizon)  \leq\margin_{\operatorname{thresh}}.
\end{split}
\label{eqn:implementation-iirl}
\end{equation}
Due to the non-linear margin constraint in \eqref{eqn:implementation-iirl}, the optimization problem can be solved using a general purpose non-linear programming solver, for example, \verb|fmincon| in MATLAB, to obtain a local minimum.
\end{algorithm}

\section{How to Mask Cognition from  Detector?} \label{sec:stoch_irl}

The framework considered in
 Theorem~\ref{thrm:irp} was deterministic; we assumed that the adversary had accurate measurements of the radar's responses. In this section, we generalize Theorem~\ref{thrm:irp} to the case where the adversary has {\em noisy} measurements of the radar's decisions. That is, the noise term $\resnoise_\time$ in the radar's response measurement $\nresponse_\time$ in \eqref{eq:model} of Definition~\ref{def:radar-adversary-interaction} is a non-zero random variable with pdf $f_\resnoise$. 
If the adversary deploys a Neyman-Pearson\footnote{
By Neyman Pearson's lemma~\cite{Van68},  it is impossible to maximize the Type-I and Type-II error of a detector simultaneously. 
In this paper, we focus on mitigating the detector by maximizing its {\em conditional} Type-I error probability.} 
type detector, how can we design  our cognition masking strategy to spoof this detector so that the radar can hide its utility and constraints? Before generalizing Theorem~\ref{thrm:irp} to the noisy case, we first address the following question: {\em How do the adversary's IRL algorithms, Theorems~\ref{thrm:rp} and \ref{thrm:rp_constraint}, adapt to noisy measurements?}

\subsection{Noisy Adversarial IRL Detectors against Cognitive Radars}\label{sec:noisy_irl}
Our key IRL results for noisy radar measurements are outlined in Definition~\ref{def:noise_rp} below. Recall from Sec.\,\ref{sec:background} that the adversary's IRL algorithm in Theorem~\ref{thrm:rp} comprises a linear feasibility test to identify a feasible strategy that rationalizes the radar's responses. When the adversary has noisy measurements of the radar's response, the deterministic feasibility test generalizes to a {\em feasibility hypothesis test} to detect the existence of feasible strategies (utilities and constraints) so that the radar responses satisfy utility maximization~\eqref{eqn:abstract_cog_radar}. 

For our hypothesis tests below, let $H_0$  and $H_1$ denote the null and alternate hypotheses that the adversary's noise-less datasets defined in \eqref{eqn:dataset_IRL_utility} and \eqref{eqn:dataset_IRL_constraint} pass, and not pass, respectively, the IRL feasibility tests~\eqref{eqn:abstract_IRL_utility} and \eqref{eqn:abstract_IRL_constraint}, respectively.
\begin{equation}
\begin{split}
    \hspace{-0.3cm}H_0:&~\text{Radar is a constrained utility maximizer~\eqref{eqn:abstract_cog_radar}}\\
    \hspace{-0.3cm}H_1:&~\text{Radar is NOT a constrained utility maximizer~\eqref{eqn:abstract_cog_radar}}
\end{split}
 \label{eqn:hypotheses}
\end{equation}
The two types of error that arise in hypothesis testing are Type-I and Type-II errors. In the radar context, the Type-I and Type-II errors have the following interpretation:
\begin{equation}
\begin{split}
    \textbf{Type-I:}~&\text{Classify a cognitive radar as non-cognitive}\\
    \textbf{Type-II:}~&\text{Classify a non-cognitive radar as cognitive}
\end{split}
\label{eqn:hyptest_errors}
\end{equation}

In analogy to Theorems~\ref{thrm:rp} and \ref{thrm:rp_constraint}, our IRL detectors defined below assume two scenarios, namely, Assumptions~\ref{asmp:utility_noise} and \ref{asmp:constraint_noise} that generalize Assumptions~\ref{asmp:utility} and \ref{asmp:constraint}, respectively, to the case where the adversary has noisy response measurements. 
\begin{assumption}
    Consider the radar-adversary interaction scenario of Assumption~\ref{asmp:utility}. The adversary has access to the noisy dataset $\ndataset_\nonlinb$ defined as:
    \begin{equation}\label{eqn:dataset_IRL_utility_noise}
       \ndataset_\nonlinb = \{\nonlinbrad(\probe_\time,\cdot),\nresponse_\time\}_{\time=1}^\horizon,~\nresponse_\time = \response_\time+\resnoise_\time,~\resnoise_\nonlinb\sim f_\resnoise
    \end{equation}
    where $\nonlinbrad(\probe_\time,\cdot)$ is defined in \eqref{eqn:special_case_cog_radar_utility}, $\response_\time$ is the radar's response and $\resnoise_\time$ is the adversary sensor's measurement noise~\eqref{eq:model} with pdf $f_\resnoise$ known to the radar.\\
    {\em IRL objective.} The adversary uses the IRL detector~\eqref{eqn:stat_test_constraint} in Definition~\ref{def:noise_rp} to detect if the noise-free dataset $\dataset_\nonlinb$~\eqref{eqn:dataset_IRL_utility} passes the IRL test~\eqref{eqn:abstract_IRL_utility} of Theorem~\ref{thrm:rp}
\label{asmp:utility_noise}
\end{assumption}
\begin{assumption}
Consider the radar-adversary interaction scenario of Assumption~\ref{asmp:constraint}. The adversary has access to the noisy dataset $\ndataset_\utility$ defined as:
    \begin{equation}\label{eqn:dataset_IRL_constraint_noise}
       \ndataset_\utility = \{\utilityrad(\probe_\time,\cdot),\nresponse_\time\}_{\time=1}^\horizon,~\nresponse_\time = \response_\time+\resnoise_\time,~\resnoise_\nonlinb\sim f_\resnoise
    \end{equation}
    where 
    $\response_\time$ is the radar's response, $\resnoise_\time$ is the adversary sensor's measurement noise~\eqref{eq:model} with pdf $f_\resnoise$ known to the radar.\\
    {\em IRL objective.} The adversary uses the IRL detector~\eqref{eqn:stat_test_constraint} in Definition~\ref{def:noise_rp} to detect if the noise-free dataset $\dataset_\utility$~\eqref{eqn:dataset_IRL_constraint} passes the IRL test~\eqref{eqn:abstract_IRL_constraint} of Theorem~\ref{thrm:rp_constraint}
\label{asmp:constraint_noise}
\end{assumption}
Our IRL hypothesis tests for detecting radar's cognition (feasible utilities and resource constraints) for noisy radar response measurements are stated in Definition~\ref{def:noise_rp} below.

\begin{definition}[IRL Detectors for Noisy Response Measurements] 
Consider the cognitive radar~\eqref{eqn:abstract_cog_radar} from Definition~\ref{def:cognitive-radar} and the radar-adversary interaction from Definition~\ref{def:radar-adversary-interaction}.\\
1. \underline{IRL for detecting feasible utilities.} Suppose Assumption~\ref{asmp:utility_noise} holds. 
Then, the statistical test below detects if the radar's responses satisfy utility maximization behavior~\eqref{eqn:abstract_cog_radar}:
\begin{align}
&\boxed{ \prob(\phi^*_\utility(\ndataset_\nonlinb)\leq L_\nonlinb)
\lessgtr_{H_0}^{H_1} \siglevel }. \label{eqn:stat_test}
\end{align}
\noindent 2. \underline{IRL for detecting feasible resource constraints.} Suppose Assumption~\ref{asmp:constraint_noise} holds. Then, the statistical test below detects if the radar's responses satisfy utility maximization behavior~\eqref{eqn:abstract_cog_radar}:
\begin{align}
&\boxed{ \prob(\phi^*_\nonlinb(\ndataset_u)\leq L_\utility)
\lessgtr_{H_0}^{H_1} \siglevel }. \label{eqn:stat_test_constraint}
\end{align}
In the statistical tests~\eqref{eqn:stat_test} and \eqref{eqn:stat_test_constraint}:
\begin{compactitem}
\item $\gamma\in[0,1]$ is the `significance level' of the test.
\item $L_\nonlinb$ and $L_\utility$ are random variables defined as:
\begin{align}
        L_\nonlinb \equiv \max_{s,\time} & 
~\probe_\time'(\resnoise_\time-\resnoise_s)\label{eqn:noise_statistic}\\
L_\utility \equiv \max_{s,\time}& 
    ~(\utilityrad(\probe_\time,\nresponse_\time) - \utilityrad(\probe_\time,\nresponse_s) )~ - \nonumber\\
    &~(
    \utilityrad(\probe_\time,\nresponse_\time-\resnoise_\time))    - \utilityrad(\probe_\time,\nresponse_s-\resnoise_s)), \label{eqn:noise_statistic_constraint}
\end{align}
where $\resnoise_\time\sim f_\resnoise$ is the measurement noise in the adversary's measurement of the radar's response~\eqref{eq:model}.
\item The test statistics $\phi^{\ast}_\nonlinb(\cdot)$ and $\phi^{\ast}_\utility(\cdot)$ are the minimum perturbations required for the noisy datasets $\ndataset_g$ and $\ndataset_\utility$, respectively, to pass the IRL feasibility tests~\eqref{eqn:abstract_IRL_utility} and \eqref{eqn:abstract_IRL_constraint}:
\begin{align}
    \phi^{\ast}_\utility(\ndataset_\nonlinb) & = \min_{\epsilon,\param>0}\epsilon,~\AFT(\param,\ndataset_\nonlinb +\epsilon) \leq 0,\label{eqn:def_suffstat}\\
    \phi^{\ast}_\nonlinb(\ndataset_u)  & = \max_{\epsilon,\paramcons>0} \epsilon,~\AFT(\paramcons,\ndataset_\utility - \epsilon) \geq 0,\label{eqn:def_suffstat_constraint}
\end{align}
\end{compactitem}
\label{def:noise_rp}
\end{definition}
\noindent{\em Remarks.} 1. The random variable $L_\nonlinb$ \eqref{eqn:noise_statistic} bounds the perturbation needed for $\ndataset_\nonlinb$ to pass the IRL test~\eqref{eqn:abstract_IRL_utility}, if $H_0$ holds:
\begin{equation*}
\hspace{-0.25cm} H_0:~\exists \param>0\text{ s.t. }\AFT(\param,\dataset_\nonlinb)\leq 0\implies \AFT(\param,\ndataset_\nonlinb + L_\nonlinb)\leq 0,
\end{equation*}
where $\dataset_\nonlinb$ is the noise-free version of the noisy dataset $\ndataset_\nonlinb$. Similarly, the random variable $L_\utility$ \eqref{eqn:noise_statistic_constraint} bounds the perturbation needed for $\ndataset_\utility$ to pass the IRL test~\eqref{eqn:abstract_IRL_constraint}, if $H_0$ holds:
\begin{equation*}
\hspace{-0.25cm} H_0:~\exists \param>0\text{ s.t. }\AFT(\param,\dataset_\utility)\geq 0\implies \AFT(\param,\ndataset_\utility + L_\utility)\geq 0,
\end{equation*}
where $\dataset_\utility$ is the noise-free version of the noisy dataset $\ndataset_\utility$.\\
\noindent 2. The IRL detectors~\eqref{eqn:stat_test} and \eqref{eqn:stat_test_constraint} classify the radar as a utility maximizer if the perturbation needed for the feasibility of the IRL inequalities lies under a particular threshold, and vice versa. Consider the statistical test of \eqref{eqn:stat_test}. Eq.\,\ref{eqn:stat_test} can be expressed differently as:
\begin{equation}\label{eqn:stat-test-alternative-representation}
    \phi_\utility^\ast(\ndataset_\nonlinb) \lessgtr_{H_1}^{H_0} F_{L_\probe}^{-1}(1-\siglevel),
\end{equation}
where the RHS term in \eqref{eqn:stat-test-alternative-representation} is the test threshold for test statistic $\phi^\ast_\utility(\cdot)$. Intuitively, the larger the perturbation needed for the feasibility of the IRL inequalities, the less confidence the adversary has to classify the radar as a utility maximizer.

{\em Computational Complexity of IRL Detectors.} The constrained optimization problems \eqref{eqn:def_suffstat} and \eqref{eqn:def_suffstat_constraint} are non-convex since the RHS of the constraint is bilinear in the feasible variable. However, since the objective function depends only on a scalar, a 1-dim.\ line search algorithm can be used to solve for $\phi^{\ast}_{\utility}(\cdot)$ in \eqref{eqn:def_suffstat} and $\phi^{\ast}_{\nonlinb}(\cdot)$ in \eqref{eqn:def_suffstat_constraint}. That is, for any fixed value of $\epsilon$, the constraints in \eqref{eqn:def_suffstat},~\eqref{eqn:def_suffstat_constraint} specialize to a set of linear inequalities for which feasibility is straightforward to check.

We now discuss a key feature of the statistical tests~\eqref{eqn:stat_test} and \eqref{eqn:stat_test_constraint} in Theorem~\ref{thrm:noise_irl_bound} that bounds the Type-I error probability $\prob(H_1|H_0)$ of the IRL detectors. Recall that the Type-I error probability is the probability of incorrectly classifying the radar as non-cognitive, when the radar's response is the solution of a constrained utility maximization problem~\eqref{eqn:abstract_cog_radar}.

\begin{theorem}[Performance of IRL Detectors (Definition~\ref{def:noise_rp})] Consider the statistical tests~\eqref{eqn:stat_test} and \eqref{eqn:stat_test_constraint} in Definition~\ref{def:noise_rp}. The Type-I error probability of the tests is bounded by the significance level of the tests $\siglevel$:
\begin{equation}
\prob(H_1|H_0)
\leq \siglevel \text{ for both detectors \eqref{eqn:stat_test} and \eqref{eqn:stat_test_constraint}}.
\label{eqn:stat_test_bound}
\end{equation}
\label{thrm:noise_irl_bound}
\end{theorem}
The proof of Theorem~\ref{thrm:noise_irl_bound} is in Appendix~\ref{appdx:IRL_performance_bound}. The key idea in the proof is to show that, given that the null hypothesis $H_0$ holds, the random variables $L_\nonlinb$ and $L_\utility$ dominate the test statistics $\phi^\ast_\nonlinb(\ndataset_\utility)$ and $\phi^\ast_\utility(\ndataset_\nonlinb)$, respectively. Since the IRL detectors have a bounded Type-I probability, our cognition masking rationale for the noisy case discussed below is to maximize their conditional Type-I error probability. 

\subsection{Masking Cognition from IRL Detectors}\label{sec:noisy_iirl}
In the previous section, we generalize the IRL results of Theorem~\ref{thrm:rp} and \ref{thrm:rp_constraint} in Sec.\,\ref{sec:background} to the case where the adversary has noisy measurements of the radar's responses. The key idea is that the IRL feasibility tests \eqref{eqn:abstract_IRL_utility} and \eqref{eqn:abstract_IRL_constraint} generalize to IRL detectors~\eqref{eqn:stat_test} and \eqref{eqn:stat_test_constraint} in Definition~\ref{def:noise_rp}, respectively, that detect utility maximization behavior. This section addresses cognition masking when the adversary uses the IRL detectors of Definition~\ref{def:noise_rp}: {\em How to mitigate the statistical tests of \eqref{eqn:stat_test} and \eqref{eqn:stat_test_constraint} and make utility maximization detection difficult?}

{\em Intuition for hiding cognition from IRL Detectors.} 
Suppose the radar follows the cognition masking scheme of Theorem~\ref{thrm:irp} for the noisy case.
Indeed, the radar's strategy is hidden from the IRL feasibility tests of Theorems~\ref{thrm:rp} and \ref{thrm:rp_constraint}, but does not affect the performance of the IRL detectors of Definition~\ref{def:noise_rp}. To do so, the radar maximizes the {\em conditional Type-I error probability}\footnote{
The radar can at best maximize the conditional Type-I error probability to mitigate the IRL detectors as the Type-I error probability is bounded by the detectors' significance level $\siglevel$ due to Theorem~\ref{thrm:noise_irl_bound}.
} of the IRL detectors by deliberately deviates from its naive responses~\eqref{eqn:abstract_cog_radar}. The conditional Type-I error probability can be viewed as the noisy analog of the inverse of the feasibility margin in the noise-less case. 

\begin{definition}[Conditional Type-I error probability for IRL Detectors (Definition~\ref{def:noise_rp})]\label{def:cond_error_prob}
Consider the datasets $\dataset_\nonlinb$ and $\dataset_\utility$ defined in \eqref{eqn:dataset_IRL_utility} and \eqref{eqn:dataset_IRL_constraint}, and their corresponding noisy versions $\ndataset_\nonlinb$ and  $\ndataset_\utility$ defined in \eqref{eqn:dataset_IRL_utility_noise} and \eqref{eqn:dataset_IRL_constraint_noise}, respectively. Let $\phi_\utility(\ndataset_\nonlinb,\utilityrad)$ and $\phi_\nonlinb(\ndataset_\utility,\nonlinbrad)$ denote the minimum perturbations required for the tuples $(\ndataset_g,\utilityrad)$ and $(\ndataset_\utility,\nonlinbrad)$, respectively, to pass the IRL feasibility tests~\eqref{eqn:abstract_IRL_utility},~\eqref{eqn:abstract_IRL_constraint}:
\begin{equation}\label{eqn:def_suff_stat_conditional}
\begin{split}
\phi^\ast_\utility(\ndataset_\nonlinb,\utilityrad) & = \min_{\eps \geq 0} \eps,~\AFT(\utilityrad,\ndataset_\nonlinb + \eps) \leq 0 \\
    \phi^\ast_\nonlinb(\ndataset_\utility,\nonlinbrad) & = \min_{\eps \geq 0} \eps,~\AFT(\nonlinbrad,\ndataset_\utility - \eps) \geq 0, 
\end{split}
\end{equation}
where $\utilityrad$ and $\nonlinbrad$ are the radar's utility and resource constraint, respectively.
Then:\\
1. For IRL detector~\eqref{eqn:stat_test}, the conditional Type-I error probability, conditioned on $\ndataset_\nonlinb$~\eqref{eqn:dataset_IRL_utility_noise} and radar's utility $\utilityrad$ is given by $\prob(H_1|\dataset_\nonlinb,\utilityrad)$, and defined as:
\begin{equation}
\label{eqn:def_cond_error_prob}
\begin{split}
\prob(H_1|\dataset_\nonlinb,\utilityrad) &= \prob(~\phi^\ast_{\utility}(\ndataset_\nonlinb,\utilityrad) > F_{L_\nonlinb}^{-1}(1-\siglevel)~)
\end{split}
\end{equation}
2. For IRL detector~\eqref{eqn:stat_test_constraint}, the conditional Type-I error probability conditioned on $\ndataset_\utility$~\eqref{eqn:dataset_IRL_constraint_noise} and radar's constraint $\nonlinbrad$ is given by $\prob(H_1|\ndataset_\utility,\nonlinbrad)$, and defined as:
\begin{equation}
\label{eqn:def_cond_error_prob_constraint}
\begin{split}
\prob(H_1|\dataset_\utility,\nonlinbrad) &= \prob(~\phi^\ast_\nonlinb(\ndataset_\utility,\nonlinbrad) > F_{L_\utility}^{-1}(1-\siglevel)~),
\end{split}
\end{equation}
In \eqref{eqn:def_cond_error_prob},~\eqref{eqn:def_cond_error_prob_constraint}, the alternate hypothesis event $H_1$ is expressed differently in the equivalent representation form of \eqref{eqn:stat-test-alternative-representation}, and the random variables $L_\utility,~L_\nonlinb$ are defined in \eqref{eqn:noise_statistic} and \eqref{eqn:noise_statistic_constraint}.
\end{definition}
\noindent {\em Remarks.}\\
1. The test statistics of the IRL detectors defined in \eqref{eqn:def_cond_error_prob} and \eqref{eqn:def_cond_error_prob_constraint} are computed via an optimization over $\reals^{2\horizon + 1}_+$, whereas the optimization in \eqref{eqn:def_cond_error_prob_constraint} is over $\reals_+$. Hence, $\phi^{\ast}_{\utility}(\ndataset_\nonlinb,\utilityrad)$ and $\phi^{\ast}_{\nonlinb}(\ndataset_\utility,\nonlinbrad)$~\eqref{eqn:def_cond_error_prob_constraint} are cheaper to compute than the test statistics $\phi^{\ast}_\utility(\ndataset_\nonlinb)$~\eqref{eqn:def_cond_error_prob} and $\phi^{\ast}_\nonlinb(\ndataset_\utility)$~\eqref{eqn:def_cond_error_prob_constraint}, respectively.\\
2. The IRL detector performance is already constrained due to Theorem~\ref{thrm:noise_irl_bound} (bounded Type-I error probability). Hence, to mitigate the IRL detector, the best the radar can do is to maximize its conditional Type-I error probability using the statistics defined in \eqref{eqn:def_suff_stat_conditional}.

We are now ready to state our cognition masking result, Theorem~\ref{thrm:noise_irp}, that mitigates IRL detectors (Definition~\ref{def:noise_rp}). In analogy to Theorem~\ref{thrm:irp} for mitigating the IRL feasibility tests of Theorems~\ref{thrm:rp} and \ref{thrm:rp_constraint}, 
the radar deliberately degrades it performance to maximize the IRL detectors' conditional Type-I error probability defined in \eqref{eqn:def_cond_error_prob} and \eqref{eqn:def_cond_error_prob_constraint}.

\begin{theorem}[Masking Cognition from Adversarial IRL Detectors] \label{thrm:noise_irp} 
Consider the cognitive radar~\eqref{eqn:abstract_cog_radar} from Definition~\ref{def:cognitive-radar}. Let $\{\response_\time^\ast\}_{\time=1}^\horizon$ denote the naive response sequence~\eqref{eqn:abstract_cog_radar} that maximizes the cognitive radar's utility. Then:\\
1. \underline{Masking Utility Function from Detector.} Suppose Assumption~\ref{asmp:utility_noise} holds. Then, the response sequence defined below makes cognition detection difficult by ensuring the detector~\eqref{eqn:stat_test} has a sufficiently large conditional Type-I error probability:
\begin{align}
   \hspace{-0.4cm} \{\pertresponse^\ast_{1:\horizon}\} = & \underset{\{\response_\time\geq \mathbf{0},~\probe_\time'\response_\time\leq 1\}}{\argmin}~
   \sum_{\time=1}^\horizon \utilityrad(\response_\time)-\utilityrad(\idresponse_\time) 
   - \lambda~\prob(H_1|\dataset_\nonlinb,\utilityrad) \label{eqn:noise_irp}
\end{align}
\noindent 2. \underline{Masking Resource Constraint from Detector.} Suppose Assumption~\ref{asmp:constraint_noise} holds. Then, the response sequence 
below makes cognition detection difficult by ensuring the detector~\eqref{eqn:stat_test_constraint} has a sufficiently large conditional Type-I error probability:
\begin{align}
  \hspace{-0.5cm}\{\pertresponse^\ast_{1:\horizon}\} = & \underset{\{\response_\time\geq \mathbf{0},~\nonlinbrad(\response_\time)\leq \thresh_\time\}
  }{\argmin}~
   \sum_{\time=1}^\horizon \utilityrad(\response_\time)-\utilityrad(\idresponse_\time) 
   - \lambda~\prob(H_1|\dataset_\utility,\nonlinbrad) \label{eqn:noise_irp_constraint}
\end{align}
In \eqref{eqn:noise_irp},~\eqref{eqn:noise_irp_constraint}, 
the positive scalar $\lambda$ parametrizes the extent of mitigation of the IRL detector.
\end{theorem}
Theorem~\ref{thrm:noise_irp} is our second result for cognition masking; see Algorithm~\ref{alg:noise_irp_utility} for a step-wise procedure for masking cognition in noise~\eqref{eqn:noise_irp} when the adversary knows the radar's constraints. Eq.\,\ref{eqn:noise_irp} and \ref{eqn:noise_irp_constraint} compute the {\em optimal} sub-optimal radar response that sufficiently hides the radar's cognition from being detected by the IRL hypothesis tests of Definition~\ref{def:noise_rp}. The parameter $\lambda$ in Theorem~\ref{thrm:noise_irp} is analogous to parameter $\eta$ in Theorem~\ref{thrm:irp}. A larger value of $\lambda$~\eqref{eqn:noise_irp} results in a larger conditional Type-I error probability for the IRL detector (larger adversarial confusion) while increasing the radar's deviation from its optimal response (greater performance degradation), and vice versa. 

The optimization problems~\eqref{eqn:noise_irp} and \eqref{eqn:noise_irp_constraint} can be solved by stochastic gradient algorithms. Algorithm~\ref{alg:noise_irp_utility} outlines a constrained SPSA~\cite{Spa03,WA99} implementation for computing the cognition masking scheme of Theorem~\ref{thrm:noise_irp}. The objective function is non-convex in the radar's responses, hence SPSA converges to a local optimum. SPSA is a generalization of adaptive algorithms where the gradient computation in \eqref{eqn:noise_irp} requires only two empirical estimates of the objective function per iteration, i.e.\,, the number of evaluations is independent of the dimension of the radar's response. For decreasing step size $\eta=1/i$ \eqref{eqn:SPSA_update}, the SPSA algorithm converges with probability one to a local stationary point. For constant step size $\eta$, it converges weakly (in probability).

{\em Summary:} In this section, we generalized our cognition masking results of Theorem~\ref{thrm:irp} to the case where the adversary has noisy measurements of the radar's responses. We first generalized our adversarial IRL feasibility tests of Theorems~\ref{thrm:rp},~\ref{thrm:rp_constraint} to IRL hypothesis tests~\eqref{eqn:stat_test} and \eqref{eqn:stat_test_constraint} in Definition~\ref{def:noise_rp} to detect utility maximization behavior given noisy radar response measurements. We then  present Theorem~\ref{thrm:noise_irp} that masks the radar's cognition by making utility maximization detection erroneous by maximizing the conditional Type-I error probability of the IRL detectors via purposeful sub-optimal responses. Our cognition masking results can be extended WLOG to any sub-optimal IRL algorithm and discussed in Appendix~\ref{appdx:arb-irl}.

\section{Numerical Results for I-IRL}\label{sec:numerical_results}
In this section, we illustrate how our cognition masking results of Theorems~\ref{thrm:irp} and \ref{thrm:noise_irp} successfully confuse adversarial IRL via the two radar tracking functionalities, namely, waveform adaptation and beam allocation discussed in Sec.\,\ref{sec:background}.

\subsection{Cognition Masking via Theorem~\ref{thrm:irp} for Noise-less Adversary Measurements}
Consider the scenario where the adversary has accurate measurements of the radar's responses. Recall from Sec.\,\ref{sec:radar_tracking_examples}
that the adversary knows the radar's constraints for waveform adaptation and the radar's utility function for beam allocation. For waveform adaptation, the probe signal parametrizes the state covariance matrix of the radar's Kalman filter due to the adversary's maneuvers, and the response signal parametrizes the sensory accuracy chosen by the radar. Recall  that the probe signal $\probe_\time$ is the diagonal of the state noise covariance matrix: $\statenoisecov_\time=\operatorname{diag}[\probe_\time(1),\probe_\time(2)]$. For beam allocation, the $i^\text{th}$ component of the probe signal $\probe_\time(i)$ is the asymptotic precision of the radar tracker for target $i$. Probe $\probe_\time$ parametrizes the radar's Cobb-Douglas utility for beam allocation. Our simulation parameters for our numerical experiments are listed below in \eqref{eqn:param-numerical}. 
 
\begin{equation}\begin{split}
    &\textit{Parameters for Numerical Experiments.} \\
    \bullet~&\text{Time Horizon}~ \horizon= 50\\
    \bullet~&\text{Probe and Response dimension}~ \probedim= 4\\
    (a)~&\underline{\text{Waveform Adaptation:}}\\
    \bullet~&\text{Probe }\probe_\time(i) \overset{\text{i.i.d}}{\sim} \operatorname{Unif}(0.2,2.5),~i=1,2,\ldots,\probedim,\\
    &~\quad\quad \time=1,2,\ldots,\horizon\\
    \bullet~&\text{Utility }\utilityrad_1(\response) = \sum_{i=1}^\probedim \sqrt{\response(i)},~\utilityrad_2(\response) = \sum_{i=1}^\probedim \response(i)^2\\ 
    \bullet~&\text{Constraint }\nonlinbrad(\probe_\time,\response) = \probe_\time'\response-1\\
    (b)~&\underline{\text{Beam Allocation:}}\\
     \bullet~&\text{Probe }\probe_\time(i) \overset{\text{i.i.d}}{\sim} \operatorname{Unif}(0.1,0.7),~i=1,2,\ldots,\probedim\\
     &~\quad\quad \time=1,2,\ldots,\horizon\\
    \bullet~&\text{Utility }\utilityrad(\probe_\time,\response) = \underset{i=1,2,\ldots,\probedim}{\Pi} \response(i)^{\probe_\time(i)}\\
    \bullet~&\text{Constraint }\nonlinbrad(\probe_\time,\response)= \|\response\|_\paramg-\thresh_\time,\\
    &~\quad\quad\kappa = 2,~\thresh_\time \overset{\text{i.i.d}}{\sim} \operatorname{Unif}(0.5,2),~\time=1,2,\ldots,\horizon\\
\end{split}\label{eqn:param-numerical}
\end{equation}
In \eqref{eqn:param-numerical}, $\operatorname{Unif}(a,b)$ denotes uniform pdf with support $(a,b)$. The elements of the probes $\probe_\time$~\eqref{eqn:abstract_cog_radar}, and intensity thresholds $\thresh_\time$~\eqref{eqn:abstract_beam} are generated randomly and independently over time $\time=1,2,\ldots,\horizon$. For waveform adaptation, we conduct our numerical experiments for two distinct utility functions $\utilityrad_1$ and $\utilityrad_2$~\eqref{eqn:param-numerical}. Given the probe sequence $\{\probe_\time\}_{\time=1}^\horizon$, the cognitive radar chooses its smart response sequence via \eqref{eqn:irp} for masking optimal waveform adaptation, and via \eqref{eqn:irp_constraint} for masking optimal beam allocation. Recall from Sec.\,\ref{sec:background} that response $\response_\time$ is the diagonal of the inverse of radar's observation noise covariance matrix for waveform adaptation. For beam allocation, $\response_\time(i)$ is the beam intensity directed towards target $i$ at time $\time$.

\begin{figure}[ht]
    \centering
  \includegraphics[width=0.8\linewidth]{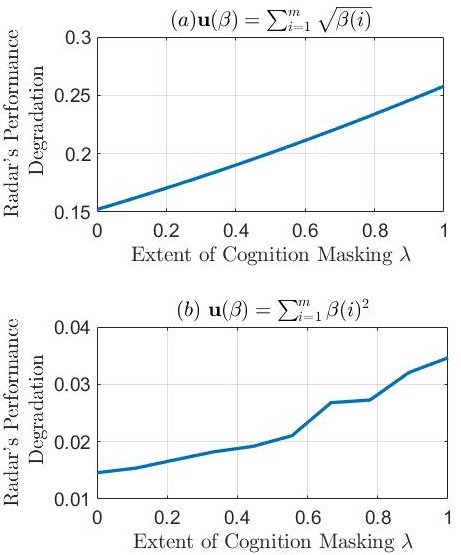}
    \caption{Masking Waveform Adaptation Strategy from Adversarial IRL. Small deliberate performance loss (vertical axis) of the cognitive radar results in large performance mitigation of the adversary (horizontal axis). The figure illustrates a cognitive radar operating with two distinct utility functions.\\
    (i) $\eta=1$ corresponds to maximum cognition masking and hence results in maximum performance loss. (ii) For a fixed value of $\eta$, the quadratic utility (sub-figure (b)) requires smaller perturbation ($\approx$ 10 times) from the optimal response 
    compared to the sub-linear utility of sub-figure (a).}
    \label{fig:numerical_waveform}
\end{figure}

\begin{figure}
    \centering
    \includegraphics[width = 0.85\columnwidth]{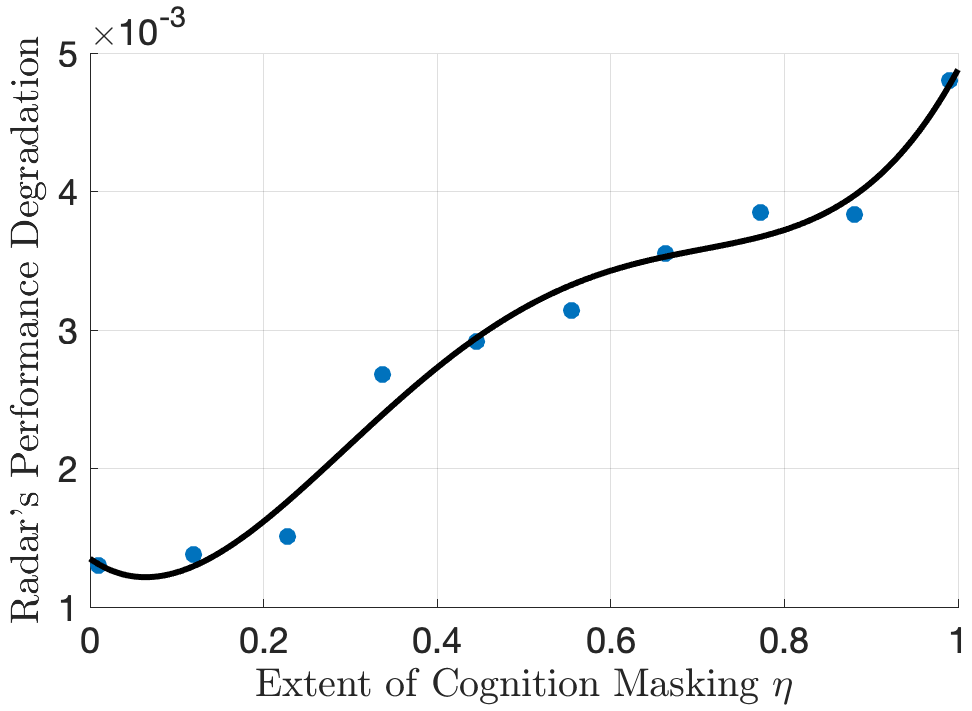}
    \caption{Masking Beam Allocation Strategy from Adversarial IRL: Small deliberate utility loss of the radar (vertical axis) results in large performance loss (extent of strategy masking $\eta$) of the adversarial IRL algorithm (horizontal axis). $\eta=0$ corresponds to zero strategy masking, and $\eta=1$ corresponds to complete strategy masking by the decision maker. As expected, the deliberate utility loss of the radar increases with $\eta$.
    }
    \label{fig:numerical_beam}
\end{figure}

Figures~\ref{fig:numerical_waveform} and \ref{fig:numerical_beam}  show the performance loss (minimum perturbation from optimal response computed via \eqref{eqn:irp},~\eqref{eqn:irp_constraint} in Theorem~\ref{thrm:irp}) of the cognitive radar as a function of $\eta$ (extent of cognition masking) when the cognitive radar performs waveform adaptation and beam allocation, respectively. We see that for both functionalities, both the radar's performance loss and adversarial IRL mitigation increase with $\eta$. This is expected since larger $\eta$ implies a larger shift of the set of feasible strategies computed via IRL to ensure the radar's strategy is sufficiently close to the edge of the feasible set, at the cost of greater deviation from the radar's optimal strategy.

\subsection{Cognition Masking via Theorem~\ref{thrm:noise_irp} for Noisy Adversary Measurements}
We now consider the scenario where the adversary has noisy measurements of the radar's response. Consider the simulation parameters of \eqref{eqn:param-numerical}. For our second set of numerical experiments for both waveform adaptation and beam allocation, we set the noise pdf $f_{\resnoise}$~\eqref{eq:model} to $\normal(0,0.3I)$, where $\normal(\mu,\Sigma)$ denotes the multivariate normal distribution with mean $\mu$ and covariance $\Sigma$, and $I$ denotes the identity matrix. In Theorem~\ref{thrm:noise_irp}.

For the noisy case, we consider only a single utility function for waveform adaption, namely, $\utilityrad(\response)=\sum_{i=1}^\probedim \sqrt{\response(i)}$. We performed our numerical experiments for three values of $\siglevel=\{0.05,0.1,0.2\}$ for both waveform adaptation and beam allocation. Recall from Sec.\,\ref{sec:stoch_irl} that $\alpha$ is the significance level of the adversary's IRL detectors \eqref{eqn:stat_test} and \eqref{eqn:stat_test_constraint} in Definition~\ref{def:noise_rp}.

Given the probe sequence $\{\probe_\time\}_{\time=1}^\horizon$, we generated the cognition masking response sequence via \eqref{eqn:noise_irp} for waveform adaption and \eqref{eqn:noise_irp_constraint} for beam allocation by varying the parameter $\lambda$~\eqref{eqn:noise_irp} over the interval $[10^0,10^5]$. Recall from Theorem~\ref{thrm:noise_irp} that the radar minimizes the detectors' conditional Type-I error probability~\eqref{eqn:def_cond_error_prob},~\eqref{eqn:def_cond_error_prob_constraint} to mitigate adversarial IRL while deliberately compromising on its performance (utility).

\begin{figure*}[ht]
    \centering
    \includegraphics[width = 0.8\linewidth]{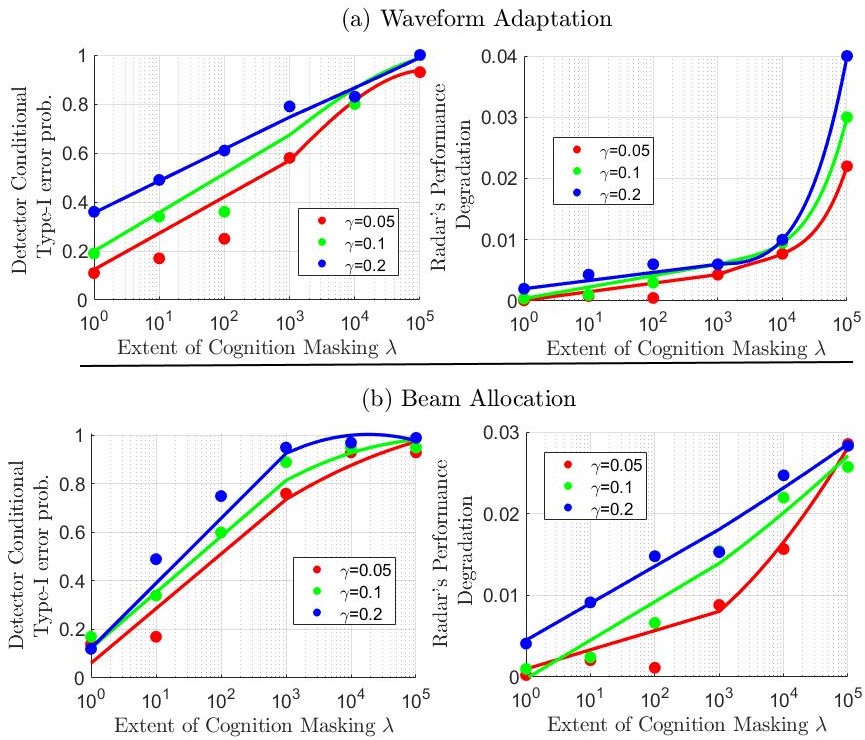}
    \caption{Masking Cognition from IRL Detectors. Performance of meta-cognitive radar for waveform adaptation (sub-figure (a)) and beam allocation (sub-figure (b)) when the adversary deploys an IRL detector~\eqref{eqn:stat_test},~\eqref{eqn:stat_test_constraint} for cognition detection. The key takeaway is that a small sacrifice in performance of the radar results in large performance loss of adversary's IRL detector. The performance loss of both the radar and the adversary due to meta-cognition increase with scaling factor $\lambda$~\eqref{eqn:noise_irp} and significance level $\siglevel$ of the adversary's IRL detectors \eqref{eqn:stat_test},~\eqref{eqn:stat_test_constraint}.}
    \label{fig:numerical_waveform_noise}
\end{figure*}

Our SPSA algorithm~\cite{Spa03,WA99} (Algorithm~\ref{alg:noise_irp_utility}) for stochastic gradient descent was executed over $10^4$ iterations for all pairs of $(\lambda,\siglevel)$, $\lambda\in\{10^0,10^1,10^2,10^3,10^4,10^5\}$ and $\siglevel\in\{0.05,0.1,0.2\}$. Figure~\ref{fig:numerical_waveform_noise} shows the conditional Type-I error probability (adversarial IRL mitigation) of the detector and performance loss of the radar as the parameter $\lambda$ is varied for three different values of the significance level $\alpha$ of the adversary's detector. Recall from Theorem~\ref{thrm:noise_irp} that the parameter $\lambda$ controls the extent of cognition masking for noisy inverse IRL. From Fig.\,\ref{fig:numerical_waveform_noise}, we see that both 
the conditional Type-I error probability of the IRL detectors and radar's performance loss increase with $\lambda$ as well as $\siglevel$. 

If $\lambda=0$, the radar simply transmits its naive response that maximizes its utility (no performance loss) and also results in zero adversarial mitigation.
For the limiting case of $\lambda\to\infty$, the radar's cognition masking response computed via Theorem~\ref{thrm:noise_irp} degenerates to a constant for all time $\time$, hence maximizing the conditional Type-I error probability of the detector at the cost of maximal performance loss for the radar.

Let us briefly discuss the variation of the radar performance and adversarial mitigation as the parameter $\siglevel$ is varied. $\siglevel$~\eqref{eqn:stat_test} can be viewed as the risk-aversion tendency of the adversary's IRL system since it bounds the detector's Type-I error probability. Recall from \eqref{eqn:hypotheses} that the Type-I error is the probability of detecting a cognitive radar as non-cognitive. Higher $\siglevel$ implies the detector is {\em risk-seeking} and a lower $\siglevel$ implies the detector is {\em risk-averse}. Naturally, a larger deviation from optimal strategy is required to mitigate a risk-averse detector compared to a risk-seeking detector by the same amount.

\begin{algorithm} \caption{SPSA for Mitigating Utility Maximization Detection for Adversarial IRL Detector~\eqref{eqn:stat_test}~(\eqref{eqn:noise_irp} in Theorem~\ref{thrm:noise_irp})} \label{alg:noise_irp_utility}
Step 1. Set $\vecresponse_0 = \{\idresponse_{1:\horizon}\}$, the naive response sequence~\eqref{eqn:abstract_waveform} that maximizes the radar's utility~\eqref{eqn:abstract_cog_radar}. \\
Step 2. Choose $\lambda>0$ (extent of cognition masking).\\
Step 3. For iterations $\iter=0,1,2,\ldots$,\\
(i) Compute $\empprob(H_1|\{\probe_\time\}_{\time=1}^\horizon,\vecresponse_i,\utilityrad)$, the empirical probability estimate of the conditional Type-I error probability of the detector~\eqref{eqn:stat_test} defined in
~\eqref{eqn:def_cond_error_prob} using $R\times\horizon$ fixed realizations  $\{\resnoise_{r,\time}\}_{r,\time=1}^{R,\horizon}$ of adversary's measurement noise~$\resnoise_\time\sim f_\resnoise$~\eqref{eq:model}:
\begin{equation}\label{eqn:emp_error_prob}
    \hspace{-0.3cm}\frac{1}{R}\sum_{r=1}^{R}\mathbbm{1}\left\{\phi_\utility^\ast(\{\probe_\time,\response_{\iter,\time}+
 \resnoise_{r,\time}\}_{\time=1}^\horizon,\utilityrad) > F_{L_\nonlinb}^{-1}(1-\siglevel)\right\}
\end{equation}
In \eqref{eqn:emp_error_prob}:\\
$\bullet$ $\vecresponse_i\equiv\{\response_{i,1:\horizon}\}\geq\mathbf{0}$ is a vector of responses\\
$\bullet$ $\mathbbm{1}\{\cdot\}$ denotes the indicator function\\
$\bullet$ $R$ controls the accuracy of the empirical probability estimate\\
$\bullet$ $F_{L_{\nonlinb}}(\cdot)$ is the distribution function of the r.v.\, $L_\nonlinb$ \eqref{eqn:stat_test}\\
$\bullet$ The statistic $\phi^\ast_\utility(\cdot,\utilityrad)$ is defined in \eqref{eqn:def_suff_stat_conditional}.\vspace{0.15cm}\\
(ii) Compute empirical estimate $\widehat{J}(\vecresponse_i)$ of objective $J(\vecresponse_i)$:
\begin{equation}\label{eqn:emp_objcompute}
\hspace{-0.5cm}\widehat{J}(\vecresponse_\iter)  = \sum_{\time=1}^\horizon \utilityrad(\idresponse_\time) - \utilityrad(\response_{\iter,\time}) -\lambda~\empprob(H_1|\{\probe_\time\}_{\time=1}^\horizon,\vecresponse_i,\utilityrad),
\end{equation}
where $\empprob(H_1|\{\probe_\time\}_{\time=1}^\horizon,\vecresponse_i,\utilityrad)$ is computed in \eqref{eqn:emp_error_prob}.\vspace{0.15cm}\\
(ii) Compute the estimate of the gradient $\nabla_{\vecresponse}~J(\vecresponse_\iter)$ as:
\begin{equation*}\label{eqn:compute_empgrad}
    \widehat{\nabla}_{\vecresponse}(\widehat{J}(\vecresponse_\iter)) = \frac{\Delta_\iter}{\omega\|\Delta_\iter\|_F^2}~\widehat{J}(\vecresponse_\iter + \gradstep~\Delta_\iter) - \widehat{J}(\vecresponse_\iter - \gradstep~\Delta_\iter),
\end{equation*}
where $\gradstep$ is the gradient step size, $||\cdot||_2$ denotes the Frobenius norm, and $\Delta_\iter\in\{-1,+1\}^{\probedim\times\horizon}$ is a random perturbation vector whose each element is $\pm 1$ with probability $1/2$. \vspace{0.2cm} \\
(iii) Update the radar's response as:
\begin{equation}\label{eqn:SPSA_update}
 \vecresponse_{\iter+1} = \operatorname{Proj}_{S_{\probe}}\left(\vecresponse_\iter + \eta~\frac{\Delta_\iter}{\|\Delta_\iter\|_F}\widehat{\nabla}_{\vecresponse} \widehat{J}(\vecresponse_\iter)\right),
\end{equation}
where $\eta$ is the response update step size and $\operatorname{Proj}_{S_\probe}$ is the projection operator to the hyperplane $S_\probe = \{\response_{1:\horizon}: \probe_\time'\response_\time=1,\response_\time\geq \mathbf{0}\}$.\vspace{0.2cm}\\
Step 4. Set $\iter\leftarrow \iter+1$ and go to Step 3.
\end{algorithm}

\section{Conclusion and Extensions}

This paper investigated how a cognitive radar can hide its cognition from an adversary, when the adversary performs 
inverse reinforcement learning (IRL)  to estimate the radar's  utility function by observing its actions.  
The adversary's IRL estimate of the radar's strategy 
is a polytope of feasible solutions to a set of convex inequalities. 
Our first cognition masking result is Theorem~\ref{thrm:irp}. When the adversary has accurate measurements of the radar's response, cognition masking via Theorem~\ref{thrm:irp} ensures the radar's true strategy lies close to the edge of the feasibility polytope computed via adversarial IRL (true strategy poorly rationalizes adversary's dataset). When the adversary has noisy measurements of the radar's response, adversarial IRL generalizes to a cognition detector defined in Definition~\ref{def:noise_rp}. Our second cognition masking result is Theorem~\ref{thrm:noise_irp}. The key idea is to maximize the probability of the radar being classified as non-cognitive by the detector subject to a bound on the radar's performance loss. Finally, in Sec.\,\ref{sec:numerical_results}, we illustrate our cognition masking results on a cognitive radar that performs waveform adaptation and beam allocation for target tracking. We show that small purposeful deviations from the optimal strategy of the radar suffice to significantly confuse the adversarial IRL system. 

This paper builds significantly on our previous work \cite{KAEM20} on ECM for identifying cognitive radars, and~\cite{PKB22,PKB22FUSION,PKB22CDC} on ECCM for masking radar cognition. Theorem~\ref{thrm:rp_vec_g} extends IRL for cognitive radars~\cite{KAEM20} when the radar faces multiple resource constraints. The linear IRL feasibility test for a single constraint case generalizes to a mixed integer feasibility test. Theorem~\ref{thrm:irp_vec_g} generalizes the cognition masking result of \cite{PKB22} to multiple constraints.  Our previous works \cite{PKB22,PKB22FUSION,PKB22CDC} assume optimal adversarial IRL via Afriat's theorem. This paper generalizes cognition masking to sub-optimal adversarial IRL algorithms. Algorithm~\ref{alg:arb-irl} outlines a cognition scheme when the adversary uses an arbitrary IRL algorithm to estimate the radar's strategy. Theorem~\ref{thrm:misspec} provides performance bounds for our cognition masking scheme when the adversary has misspecified measurements of the radar's response. Although this paper is radar-centric, we emphasize that the problem formulation and algorithms  developed also apply to adversarial inverse reinforcement learning in general machine learning applications.

Finally, a useful extension of this paper would be to study cognition masking in a dynamic radar-adversary interaction environment in comparison to the batch-wise probe-response exchange considered in this paper. Also, how to mask cognition when the adversary knows of the radar's ECCM capability? Such an approach warrants a game-theoretic discussion in terms of a Stackelberg game where the adversary moves first and the radar responds to the adversary's probes. It is also worthwhile exploring state-of-the-art concepts in chance constrained optimization~\cite{nemirovski} and robust optimization~\cite{robust1,robust2} to achieve cognition masking under uncertainty - when the radar has noisy measurements of the adversary's probes.

\appendix
\subsection{Feasibility Test for Adversarial IRL}\label{appdx:IRL_inequalities}
\begin{definition}[IRL Feasibility Test] Consider a dataset $\dataset=\{\nonlinb_\time(\cdot),\response_\time\}_{\time=1}^\horizon$ of monotone functions $\nonlinb_\time$ and responses $\response_\time\geq\mathbf{0}$. The set of IRL feasibility inequalities $\AFT(\cdot,\dataset)$
is defined as: 
\begin{align} 
    \hspace{-0.4cm}&\quad \AFT(\param,\dataset) \leq \mathbf{0}\label{eqn:IRL_feasibility}\\
    \hspace{-0.4cm}\equiv 
    &~\param_s - \param_\time - \param_{\time + \horizon} (\nonlinb_\time(\response_s) - \nonlinb_\time(\response_\time)) \leq 0,  \nonumber \\
    &~\text{for all } s,\time\in\{1,2,\ldots,\horizon\},~s\neq\time,\label{eqn:IRL_feasibility_granular}
\end{align}    
where the feasible variable $\param\in\reals_+^{2\horizon}$.
\end{definition}
\noindent {\em Remarks.}\\
1. The IRL feasibility inequalities are linear in the feasible variable $\param$. Hence, $\AFT(\cdot,\dataset)$ is a linear feasibility test whose feasibility can be checked using a linear programming solver.\\
2. The set of inequalities $\AFT(\cdot)$ checks for relative optimality~\cite{PK20inverse} between any pair of indices $s,\time\in\{1,2,\ldots,\horizon\}$. 
Consider the abstract setup of the cognitive radar in Definition~\ref{def:cognitive-radar} and suppose Assumption~\ref{asmp:utility} holds. In this context, we define relative optimality as:
\begin{align}
    &~\text{If }\probe_\time'\response_s \leq \probe_\time'\response_\time,\text{ then }\utility(\response_\time)\geq\utility(\response_s),\label{eqn:relopt}\\
    &\text{for all }s,\time\in\{1,2,\ldots,\horizon\},~s\neq\time.\nonumber
\end{align}
Eq.\,\ref{eqn:relopt} states that $\response_\time$ is the optimal response choice from the {\em finite} set $\{\response_\time\}_{\time=1}^\horizon$ and hence a weaker notion of optimality wrt \eqref{eqn:abstract_cog_radar}. Hence, if Assumption~\ref{asmp:utility} holds, we say wrt utility $\utility(\cdot)$, the dataset $\{\probe_\time'(\cdot)-1,\response_\time\}_{\time=1}^\horizon$ satisfies relative optimality. The feasibility of the IRL inequalities \eqref{eqn:IRL_feasibility} is equivalent (due to \cite{Afr67}) to the existence of a utility function such that relative optimality holds. \\
\noindent 3. Finally, let us provide some intuition on the feasible variable $\param$ in \eqref{eqn:IRL_feasibility}. Suppose $\AFT(\param,\dataset)\leq \mathbf{0}$~\eqref{eqn:IRL_feasibility}. On inspecting \eqref{eqn:IRL_feasibility_granular} in more detail, the first $\horizon$ components $\param(1),~\param(2),\ldots,~\param(\horizon)$ of the vector $\param$ can be viewed as the utility function $\utility_\param$ corresponding to the feasible variable $\param$ evaluated at responses 
$\response_1,\response_2,\ldots,\response_\horizon$. The last $\horizon$ components $\param(\horizon+1),~\param(\horizon+2),\ldots,~\param(2\horizon)$ can be viewed as the Lagrange multipliers associated with the Lagrangian of the optimization problem $\max_{\response\geq\mathbf{0}}~\utility(\response),~\nonlinb_\time(\response)\leq 0$ over all $\time$. In summary, we have the following correspondence between the feasible variable $\param$ and the feasible utility function $\utility_\param$:
\begin{align}
    &\param(\time)  = \utility_\param(\response_\time),\nonumber\\
    &\param(\horizon+\time)  = \frac{\nabla_{\response}~\utility_\param(\response)\vert_{\response=\response_\time}}{\nabla_{\response}~\nonlinb_\time(\response)\vert_{\response=\response_\time}},~\time\in\{1,2,\ldots,\horizon\},\label{eqn:IRL_feasible_interpretation}
\end{align}
where the division operation in \eqref{eqn:IRL_feasible_interpretation} is element-wise.\\
4. The interpretation of the feasibility vector $\param$ in \eqref{eqn:IRL_feasible_interpretation} facilitates us to define the mapping $\param\rightarrow\utility_\param(\cdot)$ as:
\begin{equation}\label{eqn:utility_reconstructed}
    \utility_\param(\response) = \min_{\time\in\{1,2,\ldots,\horizon\}}~\bigg\{\param(\time) + \param(\horizon+\time)(\nonlinb_\time(\response)-\nonlinb_\time(\response_\time))\bigg\} 
\end{equation}
It is straightforward to show that both relative optimality~\eqref{eqn:relopt} and  absolute optimality \eqref{eqn:abstract_cog_radar} hold for $\utility_\param$~\eqref{eqn:utility_reconstructed} if $\AFT(\param,\dataset)\leq\mathbf{0}$. Also, for any utility function $\utility$ and dataset $\dataset$~\eqref{eqn:IRL_feasibility}, define the reverse mapping $\utility\rightarrow\utility_{\AFT}$ as:
\begin{align}
    &\utility_{\AFT}(\time)  = \utility(\response_\time),\nonumber\\
    &\utility_{\AFT}(\horizon+\time)  = \frac{\nabla_{\response}~\utility(\response)\vert_{\response=\response_\time}}{\nabla_{\response}~\nonlinb_\time(\response)\vert_{\response=\response_\time}},~\time\in\{1,2,\ldots,\horizon\}.\label{eqn:IRL_reverse_map}
\end{align}
Then, it is straightforward to show:
\begin{equation}
    \response_\time\in\argmax_{\response\geq \mathbf{0}} ~\utility(\response),~\nonlinb_\time(\response)\leq 0 \implies \AFT(\utility_{\AFT},\dataset) \leq \mathbf{0},
\end{equation}
where $\utility_{\AFT}$ is defined in \eqref{eqn:IRL_reverse_map} and can be interpreted as the finite dimensional projection of $\utility(\cdot)$ for the IRL feasibility test $\AFT(\cdot,\dataset)\leq \mathbf{0}$.

\subsection{IRL for Identifying Radar's Resource Constraint}\label{appdx:IRL_constraint}
\begin{theorem}[IRL for Identifying Resource Constraint]\label{thrm:rp_constraint}
Suppose assumption \ref{asmp:constraint} holds.
With $\dataset_\utility=\{\utility_\time(\cdot),\response_\time\}_{\time=1}^\horizon$ denoting the utility-response dataset, the adversary can identify if there exists feasible constraints that satisfies~\eqref{eqn:abstract_cog_radar} for all $\time$ by checking the feasibility of the linear inequalities $\AFT(\cdot,\dataset_\utility)$~\eqref{eqn:IRL_feasibility}:
\begin{equation}
\begin{split}
 & \exists\param\in\reals_+^{2\horizon}~\text{s.t. }\AFT(\param,\dataset_\utility)\geq \mathbf{0},\\
  \iff &\exists~\nonlinb,\thresh_\time~\text{s.t. } \response_\time \in \argmax \utility_\time(\response),~\nonlinb(\response)\leq \thresh_\time\forall \time\\
\end{split}\label{eqn:abstract_IRL_constraint}
\end{equation}
The set-valued IRL estimate of the constraint $\nonlinb$ is given by:
\begin{equation}
\begin{split}
 \nonlinb_{\IRL}(\response) & \equiv \{\nonlinb_{\IRL}(\response;\param) : \AFT(\param,\dataset_u)\geq \mathbf{0})\}\\
 \nonlinb_{\IRL}(\response;\param) & = \underset{\time\in \{1,2,\dots,\horizon\}}{\operatorname{max}}\{\param_{\time}+\param_{\time + \horizon}~(\utility_\time(\response) - \utility_\time(\response_\time))\},
\end{split}
\label{eqn:estbudget}
\end{equation}
\end{theorem}

\subsection{Example. Optimal Waveform Adaptation} \label{appdx:waveform}
Consider the radar-adversary interaction of Definition~\ref{def:radar-adversary-interaction}. Suppose the radar uses a Bayesian tracker (Kalman filter) for estimating the target's state, and suppose the adversary's probe and radar's response parameterize the radar's state and observation noise covariances, respectively. We assume
$\response_\time(i)$ is the measurement precision (amount of directed energy) of the radar in the $i^{\text{th}}$ mode, and probe $\probe_\time(i)$ is the radar's incentive for considering the $i^{\text{th}}$ mode of the target. Specifically, probe $\probe_\time$ is the vector of eigenvalues of the state noise covariance matrix $Q$, and response $\response_\time$ is the vector of eigenvalues of the inverse of the observation noise covariance matrix $R^{-1}$ in the linear Gaussian dynamics model of~\eqref{eqn:kalman-sys}. 

Our working assumption is that the radar is cognitive, and maximizes a utility $\utilityrad$ subject to an operating constraint $\nonlinbrad(\response)\leq 0$. For optimal waveform adaptation, we assume the operating constraint is a bound on the inverse of the radar tracker's asymptotic covariance $\Sigma^*(\probe_\time, \response_\time)$, where $\Sigma^*(\probe_\time, \response_\time)$ is the solution of the algebraic Ricatti equation. As justified below, a bounded asymptotic precision is equivalent to the linear constraint $\probe_\time'\response\leq 1$.

{\em Kalman filter-based Bayesian Tracker.} Suppose the radar estimates the target state $\hat{x}_n$ with covariance $\Sigma_n$ from observations $y_{1:n}$.
The posterior $\pi_n$ is propagated recursively in time via the classical Kalman filter equations:
\begin{align*}
\Sigma_{n+1|n}& = A \Sigma_n A' + Q(\probe_\time),~K_{n+1} = C \Sigma_{n+1|n} C' + R(\response_\time)\\
\psi_{n+1} = & \Sigma_{n+1|n} C' K^{-1}_{n+1}, ~\hat{x}_{n+1} = A \hat{x}_n + \psi_{n+1} (y_{n+1} - C A \hat{x}_n)\\
\Sigma_{n+1} = & (I - \psi_{n+1} C)~\Sigma_{n+1|n}.
\end{align*}
Assuming the model parameters \eqref{eqn:kalman-sys} satisfy the conditions that $[A, C]$ is detectable and $[A, \sqrt{Q}]$ is stabilizable,
the steady-state predicted covariance $\Sigma_\infty$ is the unique positive semi-definite solution of the \textit{algebraic Riccati equation} (ARE):
\begin{align}
\mathcal{A}(\probe_\time, \response_\time, \Sigma) = & 
-\Sigma + A(\Sigma -
 \Sigma C' [C \Sigma C' + R(\response)]^{-1} C \Sigma) A' \nonumber\\
 & \quad \quad + Q(\probe) = 0.\label{eqn:ARE}
\end{align}
Let $\Sigma(\probe_\time,\response_\time)^{-1}$ denote the solution of the ARE~\eqref{eqn:ARE}. 
Our working assumption is that the radar can only expend sufficient resources to ensure that the precision (inverse covariance) is at most some pre-specified precision $\Sigma_n^{* -1}$ at each epoch $n$ and then choose the `best' waveform (equivalently, observation covariance) available. Such trade-offs for the radar arise frequently in cognitive radar models in literature~\cite{cogradar_1,cogradar_2}\footnote{In \cite{cogradar_1}, the radar minimizes the mean-squared error of the target's estimate subject to a Cramer-Rao bound on the target's localization accuracy. In \cite{cogradar_2}, the radar minimizes its posterior Cramer-Rao bound, subject to an upper bound on its transmission power and antenna deployment cost. } We now invoke \cite[Lemma 3]{KAEM20} and show the equivalence between the complex non-linear constraint $(\Sigma^\ast(\probe_\time,\response_\time))^{-1}\leq \bar{\Sigma}^{-1}$ and the linear constraint $\probe_\time'\response_\time\leq 1$. The key idea in \cite[Lemma 3]{KAEM20} is to show the asymptotic precision $\Sigma^\ast(\probe_\time,\cdot)$ is monotone increasing in the second argument $\response_\time$ using the information Kalman filter formulation~\cite{AM79}. To summarize, the radar's optimal waveform adaptation strategy can be expressed as:
\begin{equation}\label{eqn:abstract-waveform-appdx}
    \boxed{\response_\time\in\argmin_{\response\geq \mathbf{0}} \utilityrad(\response),~\probe_\time'\response\leq 1 }.
\end{equation}
{\em Remark.} The RHS in the constraint in \eqref{eqn:abstract-waveform-appdx} can be set to 1 WLOG by appropriately scaling the probe signal $\probe_\time$.

\subsection{Example. Optimal Beam Allocation}\label{appdx:beam}
For abstracting beam allocation into a constrained utility maximization setup~\eqref{eqn:abstract_cog_radar}, we work at a higher level of abstraction compared to waveform adaptation. Specifically, we assume the adversary comprises multiple targets.  At this higher level of abstraction, we view each component $i$  of the adversarial probe signal $\probe_\time(i)$ as the trace of the predicted precision matrix (inverse covariance) of target $i$. Recall from the previous section that we used the probe signal~\eqref{eqn:abstract_waveform} to parametrize the maneuver covariance matrix. In comparison, we now use the trace of the precision of each target in our probe signal -- this allows us to consider multiple targets. 

{\em Multiple Target-tracking.} For the optimal beam allocation example, we assume the adversary comprises a {\em collection} of $\numtargets$ adversarial targets indexed by $i=1,2,\ldots,\numtargets$. We assume the cognitive radar adaptively switches its beam between the
$\probedim$ targets. As in (\ref{eqn:kalman-sys}), on the fast time scale indexed by $k$, target $i$ has linear Gaussian dynamics and the radar obtains linear Gaussian measurements of the targets' maneuvers:
\begin{equation} \label{eq:lineargaussian2}
\begin{split}
\state^i_{n+1} &= \statem\state^i_n  + \snoise^i_n, \quad \state_0 \sim \belief_0 \\
\obs^i_n &= \obsm\, \state^i_n + \onoise^i_n, \quad i=1,2,\ldots,\probedim
\end{split}
\end{equation}
Here $ \snoise^i_n\sim \normal(0,\snoisecov_\time(i))$,
 $\onoise_n^i \sim \normal(0,
 \onoisecov_\time(i))$.  We assume that both $\snoisecov_\time(i)$ and $\onoisecov_\time(i)$ are known to the radar and adversary. As in the previous sub-section, $\time$ indexes the slow time scale and $n$ indexes the fast time scale. For target $i$, the radar's Kalman covariance The enemy's radar tracks our $\probedim$ targets using  Kalman filter trackers.

{\em Probe-Response Parametrization.} The $i^\text{th}$ target's predicted {\em precision} parametrizes the $i^\text{th}$ element of the adversary's probe. Specifically, the price the radar pays at the start of epoch $\time$ for tracking target $i$ is the trace
of the inverse of the predicted  covariance   at  epoch $\time$ using the Kalman predictor:
\begin{equation}
\begin{split}
 \probe_\time(i) & = \trace(\Sigma^{-1}_{\time}(i)), \quad i =1 ,\ldots, \numtargets,\\
 \Sigma_\time(i) & = \underset{n\rightarrow\infty}{\lim} A^n \Sigma_{0,t}(i) A'^{n} + \sum_{l=0}^{n-1} A^{l}\snoisecov_\time(i)A'^{l},
\label{eq:probe_beam}
\end{split}
\end{equation}
where $\Sigma_{0,t}(i)$ is the covariance of the $i^\text{th}$ target's covariance at time (fast time scale) $n=0$ in epoch $t$.
Clearly, the predicted covariance $\Sigma_{\time}(i)$ \eqref{eq:probe_beam} is a deterministic function of  the maneuver covariance $\snoisecov_\time(i)$ of target $i$. The radar's response $\response_\time(i)$ to the adversary's probe is the beam intensity allocated to target $i$ in epoch $\time$. Intuitively, at the start of every epoch $\time$, the radar computes the predicted precision of the state estimate of target $i$, and then chooses the beam intensity towards target $i$ during epoch $\time$. The radar can at best compute the {\em predicted} precision for its decision (since it has no access to observations $\obs_1,\obs_2,\ldots$ at the beginning of the epoch). 

{\em Optimal beam allocation.} We assume that the radar, at epoch $\time$, faces a resource constraint $\nonlinbrad(\response)\leq \thresh_\time$ and directs beam intensities $\response(1),\response(2),\ldots,\response(\numtargets)$ towards targets $1,2,\ldots,\numtargets$, respectively. We assume the radar's resource constraint can be expressed as $\nonlinbrad=\|\response\|_k$, the $k$-norm of the radar's response. The radar's aim is to maximize the Cobb-Douglas utility of the transmitted intensities:
\begin{equation}\label{eqn:cd-util}
\utilityrad_\time(\response) = \underset{i=1,\ldots,\numtargets}{\Pi} \response(i)^{\probe_\time(i)}.
\end{equation}
The exponents for the Cobb-Douglas utility function~\eqref{eqn:cd-util}, referred to as elasticity parameters in consumer economics literature, parameterize the marginal utility per consumer good. In complete analogy, the elasticity parameters in our cognitive radar context parameterize the incentive for the radar to focus its beam towards a particular target. The economic rationale for the cognitive radar is as follows - a higher predicted precision $\probe_\time(i)$ for target $i$ implies a {\em better} state estimate, and thus a higher incentive for the radar to direct its transmission intensity towards target $i$. To summarize, the cognitive radar's beam allocation functionality can be abstracted as:

\begin{align}
    &\boxed{\response_\time =     \underset{\response}{\argmax}~ \utilityrad_\time(\response),~  \nonlinbrad(\response) \leq \thresh_\time,} \label{eqn:abstract_beam_1}\\
    & \utilityrad_\time = \underset{i=1,\ldots,\numtargets}{\Pi} \response(i)^{\probe_\time(i)}, \nonlinbrad(\response) = \|\response\|_{\paramg},~\paramg>1\nonumber
\end{align}
Eq.\,\ref{eqn:abstract_beam} abstracts the beam allocation functionality of the cognitive radar; at time $\time$ the radar maximizes a Cobb-Douglas\footnote{
The Cobb-Douglas utility function is widely used in microeconomics to model human satisfaction from buying consumer goods. In the radar context, this utility function measures the performance of the cognitive radar.
} utility $\utilityrad_\time(\response)$ (utility specified by the adversarial target) subject to a constraint on the $k$-norm of its response (beam intensity allocation). In the beam allocation case, the aim of adversarial IRL is to estimate the scalar $\paramg$ that parametrizes the radar's budget constraint $\nonlinbrad$.

In \eqref{eqn:abstract_beam}, notice how the adversary's probe parametrizes the radar's utility function instead of its budget constraint as in \eqref{eqn:abstract_waveform}. Also, observe that both the utility function and cost are monotone in the transmission intensities - higher beam intensity yields a larger utility but is also more costly. We assume that: (1) each target $i$ is equipped with a radar detector and can estimate the beam intensity $\response_\time(i)$ the enemy's radar directs towards it - this assumption facilitates the targets to carry out adversarial IRL attacks on the radar, and (2) the adversarial targets know the radar is maximizing the Cobb-Douglas utility function~\eqref{eqn:cd-util}, but does not know the radar's budget constraint $\nonlinbrad(\response)\leq \thresh_\time$ and is the adversary's IRL objective in the beam allocation context as discussed below.

\noindent {\em IRL for optimal beam allocation.} We now present Theorem~\ref{thrm:rp}, a revealed preference-based IRL algorithm for the adversary. Unlike Theorem~\ref{thrm:rp}, the adversary parametrizes (and hence, knows) the utility function of the cognitive radar, but does not know its budget constraint. Hence, the IRL objective of the adversarial target in Theorem~\ref{thrm:rp} is to estimate the radar's budget $\nonlinbrad(\cdot)$.

\subsection{Proof of Theorem~\ref{thrm:noise_irl_bound}}\label{appdx:IRL_performance_bound}
\begin{proof}
 The Type-I error probability is given by $\prob(H_1|H_0)$, that is, probability that the adversary misclassifies a utility maximizer as not a utility maximizer. Let us first consider the statistical test~\eqref{eqn:stat_test} in Definition~\ref{def:noise_rp}. Assume $H_0$ holds. Then, the Afriat inequalities~\eqref{eqn:abstract_IRL_utility} for the true dataset $\dataset_\probe$ to have a feasible solution. Let $\{u_t,\lambda_\time\}$ denote any feasible solution to Afriat's inequalities~\eqref{eqn:abstract_IRL_utility}. Then, the following inequalities result:
 \begin{align}
    &  u_s-u_t-\lambda_\time \probe_\time'(\response_s-\response_\time) \leq 0\nonumber\\
    \Leftrightarrow~&  u_s-u_\time-\lambda_\time\probe_\time'(\response_s-\nresponse_s + \nresponse_s - \response_\time+\nresponse_\time -\nresponse_\time)  \leq 0\nonumber\\
    \Leftrightarrow~&  u_s-u_\time-\lambda_\time\left(\probe_\time'(\nresponse_s-\nresponse_\time) +  \probe_\time'(\resnoise_\time-\resnoise_s)\right)\leq 0\nonumber\\
    \Leftrightarrow~& u_s-u_\time-\lambda_\time\bigg(\probe_\time'(\nresponse_s-\nresponse_\time) +  \max_{s,\time}\probe_\time'(\resnoise_\time-\resnoise_s) \bigg)\leq 0 \label{eqn:Afriat_noise}
 \end{align}
 Since the test statistic $\phi^\ast(\ndataset_\probe)$~\eqref{eqn:def_suffstat} is the  minimum perturbation needed for the feasibility of Afriat's inequalities~\eqref{eqn:abstract_IRL_utility}, \eqref{eqn:Afriat_noise} yields the following inequality:
 \begin{equation}
 \phi^\ast(\ndataset_\probe)\leq \max_{s,\time}\probe_\time'(\resnoise_\time-\resnoise_s) \equiv L_\probe\label{eqn:test_statistic_inequality}
 \end{equation}
 The Type-I error probability can now be bounded as:
\begin{align}
\prob(H_1|H_0)& = \prob(\prob(\phi^\ast(\ndataset_\probe)\leq L_\probe)\leq \siglevel )\nonumber\\
& = \prob(\phi^\ast(\ndataset_\probe)\geq F_{L_\probe}^{-1}(1-\gamma))\nonumber\\
& \leq \prob(L_\probe\geq F_{L_\probe}^{-1}(1-\gamma))~\text{from \eqref{eqn:test_statistic_inequality}}\nonumber\\
& = 1 - \prob(L_\probe \leq F_{L_\probe}^{-1}(1-\siglevel)) = 1 - (1-\siglevel) = \siglevel \nonumber
\end{align}
Hence, the Type-I error probability of the statistical test~\eqref{eqn:stat_test} is bounded by its significance level $\siglevel$. Showing the Type-I error probability of the detector~\eqref{eqn:stat_test_constraint} is bounded is identical to the steps outlined above, and thus, omitted.
\end{proof}

\subsection{Masking Radar's Utility Function for Multiple Constraints}\label{appdx:vector-valued-g}
Our IRL results for estimating the radar's utility $\utilityrad$~\eqref{eqn:abstract_cog_radar} assumes a scalar-valued budget constraint for the radar. In general, the radar faces multiple constraints, or equivalently, the constraint $\nonlinbrad(\cdot)\leq 0$ is vector-valued. {\em How to generalize Theorem~\ref{thrm:rp} to vector-valued constraints?}
In this section, we generalize our IRL algorithm~(Theorem~\ref{thrm:rp}) and cognition masking results of Theorem~\ref{thrm:irp} to vector-valued $\nonlinbrad$ when the adversary knows the radar's constraints and estimates the radar's utility $\utilityrad$ - this scenario is formalized below in assumption \ref{asmp:utility_vec_g}. Generalizing IRL for identifying a vector-valued constraint $\nonlinbrad(\cdot)\leq 0$ is non-identifiable and hence, omitted. 
\begin{assumption}
    The radar's resource constraint $\nonlinbrad(\probe_\time,\response_\time):\reals_+^{2\probedim}\rightarrow \reals_+^{\gdim}$ in \eqref{eqn:abstract_cog_radar} is vector-valued, and the radar's utility $\utilityrad(\cdot)$ is independent of $\probe_\time$:
    \begin{equation}\label{eqn:special_case_cog_radar_utility_vec_g}
    \nonlinbrad(\probe_\time,\response) \equiv \{\nonlinbrad_i(\probe_\time,\response)\}_{i=1}^\gdim,~\utilityrad(\probe_\time,\response) \equiv \utilityrad(\response),
    \end{equation}
    where $\probedim$ is the dimension of the probe/response, and $\nonlinb_i(\cdot)$ is a scalar-valued constraint.\\
    \noindent \underline{IRL objective.} The adversary aims to reconstruct the radar's utility $\utilityrad(\cdot)$ using the dataset $\dataset_\nonlinb$, where $\dataset_\nonlinb$ is defined in \eqref{eqn:dataset_IRL_utility}.
\label{asmp:utility_vec_g}
\end{assumption}
Assumption~\ref{asmp:utility_vec_g} specializes to assumption~\ref{asmp:utility} when $\nonlinbrad$ is scalar-valued and linear in both the probe and response vectors. Let us now state Theorem~\ref{thrm:rp_vec_g} for achieving IRL for vector-valued constraints when assumption~\ref{asmp:utility_vec_g} holds.
\begin{theorem}[IRL for Identifying Radar's Utility Function for Vector-Valued Constraints]\label{thrm:rp_vec_g} Consider the cognitive radar described in Definition~\ref{def:radar-adversary-interaction}. Suppose assumption \ref{asmp:utility} holds. Then:\\
(a) The adversary checks for the existence of a feasible utility function that satisfies \eqref{eqn:abstract_cog_radar} by checking the feasibility of the following set of inequalities:
\begin{align}
   &\hspace{-0.2cm}\text{There exists a feasible }\param\in\reals^{(1+\gdim)\horizon}~\text{such that:}\nonumber\\
   &\hspace{-0.2cm}(i)~~ \param_{\time} - \param_s -\sum_{i=1}^\gdim \param_{s\gdim+i}~\nonlinb_i(\probe_s,\response_\time)\leq 0,~\forall~s,\time,\label{eqn:abstract_IRL_utility_vec_g}\\   &\hspace{-0.2cm}(ii)~\param_{1:\horizon}>\mathbf{0},~\param_{\time\gdim+1:~(\time+1)\gdim} \geq \mathbf{0} \text{ and not all zeros  }\forall\time\label{eqn:not_all_zeros_vec_g}\\
  &\hspace{-0.2cm}\Leftrightarrow \exists~\utility~\text{s.t. } \response_\time \in \argmax \utility(\response),~\nonlinbrad(\probe_\time,\response_\time)\leq \mathbf{0}~\forall\time,\nonumber
\end{align}
where dataset $\dataset_\nonlinb$ is defined in \eqref{eqn:dataset_IRL_utility}.\\
\noindent (b) If \eqref{eqn:abstract_IRL_utility_vec_g} has a feasible solution, the set-valued IRL estimate of the radar's utility $\utilityrad$ is given by:
\begin{equation}
\begin{split}
&\utility_{\IRL}(\response)  \equiv \{\utility_{\IRL}(\response;\param) : \text{\eqref{eqn:abstract_IRL_utility_vec_g} is feasible}\},\\
    &\utility_{\IRL}(\response;\param)  = \underset{\time}{\operatorname{min}}\{\param_\time+\sum_{i=1}^\gdim~\param_{\time\gdim + i} ~\left(\nonlinb_i(\probe_\time,\response)-\nonlinb_i(\probe_\time,\response_\time)\right)\},
    \end{split}
    \label{eqn:estutility_vec_g}  
\end{equation}
where $\param$ is any feasible solution to the inequalities \eqref{eqn:abstract_IRL_utility_vec_g}, \eqref{eqn:not_all_zeros_vec_g}.
\end{theorem}
\noindent {\em Remarks.}\\
1. In comparison to the linear feasibility test~\eqref{eqn:abstract_IRL_utility} of Theorem~\ref{thrm:rp}, \eqref{eqn:abstract_IRL_utility_vec_g} in Theorem~\ref{thrm:rp_vec_g} is a mixed-integer linear feasibility test, mixed-integer due to the second set of inequalities~\eqref{eqn:not_all_zeros_vec_g} in the feasibility test. \\
2. Afriat's theorem~\ref{eqn:abstract_IRL_utility} requires that the constraint be active at the solution, meaning $\probe_\time'\response_\time=1$ for all $\time$. This requirement is implicitly satisfied for the scalar case due to the monotonicity of both the utility $\utilityrad(\cdot)$ and constraint $\probe_\time'(\response)\leq 1$. For vector-valued $\nonlinbrad$, however, requiring all constraints to be active at the solution is highly restrictive. That is, $\nonlinbrad_i(\probe_\time,\response_\time)=0$ for all time steps $\time$ and constraint indices $i$ is not true in general for a cognitive radar. Hence, the IRL inequalities for multiple constraints must account for the inactive constraints for all time steps $\time$. More precisely, the inverse learner needs to check for at least one active constraint out of all $\gdim$ resource constraints for all $\time$. This is ensured by the feasibility of \eqref{eqn:not_all_zeros_vec_g} in Theorem~\ref{thrm:rp_vec_g}. At a deeper level, \eqref{eqn:not_all_zeros_vec_g} tests for complementary slackness in the KKT conditions~\cite[Sec.\,5.5]{BV04} for first-order optimality of the radar's responses.\\
3. In Afriat's theorem~(Theorem~\ref{thrm:rp}), the reconstructed utility function~\eqref{eqn:estutility} is a point-wise minimum of scaled and shifted versions of the radar's linear constraints $\probe_\time'\response,~\time=1,2,\ldots,\horizon$. Intuitively, the basis functions for the adversary's estimate of the radar's utility are the radar's constraints $\{\probe_\time'\response\}_{\time=1}^\horizon$. For the multiple constraints case in Theorem~\ref{thrm:rp_vec_g} above, the adversary's utility estimate has a richer representation due to a larger set of basis functions $\{\nonlinbrad_i(\probe_\time,\response)\}_{i,\time=1}^{\gdim,\horizon}$. 

Having defined our IRL algorithm for multiple constraints in Theorem~\ref{thrm:rp_vec_g} above, we now present our cognition masking result for mitigating the IRL procedure of Theorem~\ref{thrm:rp_vec_g}. 

\begin{definition}[Feasibility Margin for Reconstructed Utility~\eqref{eqn:abstract_IRL_utility} for Multiple Constraints] \label{def:margin_utility_vec_g} 
Consider the dataset $\dataset_\nonlinb$ defined in \eqref{eqn:dataset_IRL_utility}. Suppose the radar's constraint $\nonlinbrad$ is vector-valued. The feasibility margin $\margin_u(\dataset_\nonlinb)$ defined below measures how far is the utility $\utility$ is from failing the IRL feasibility inequalities~\eqref{eqn:abstract_IRL_utility_vec_g},~\eqref{eqn:not_all_zeros_vec_g}:
\begin{equation}
\label{eqn:margin_rp_vec_g}
\begin{split}
    & \margin_{u}(\dataset_\nonlinb) =  \min_{\boldsymbol{\lambda
    }_{1:\horizon}}\Big\{\min_{s,\time}~\utility(\response_s) - \utility(\response_t) + 
\\
&\hspace{2.7cm}\left(\nabla~\nonlinbrad(\probe_s,\response_s)~\boldsymbol{\lambda}_s\right)'(\response_t-\response_s)\Big\},\\
    & \boldsymbol{\lambda}_{1:\horizon}\in\reals^\gdim,~\boldsymbol{\lambda}_\time 
    \in\argmax~G_\time(\boldsymbol{\lambda}),~\boldsymbol{\lambda}\geq 0,
\end{split}
\end{equation}
where $G_\time(\cdot)$ is the dual function of the optimization problem~\eqref{eqn:abstract_cog_radar} at time $\time$.
\end{definition}
The margin definition in \eqref{eqn:margin_rp_vec_g} above is a multi-constraint generalization of Definition~\ref{def:margin_utility}. To glean some insight into the notation in \eqref{eqn:margin_rp_vec_g} above, consider the simple case where $\gdim=1$. Then, the solution to $\argmax~G_\time(\boldsymbol{\lambda})$ is simply the Lagrange multiplier associated with the single operating constraint in the optimization problem~\eqref{eqn:abstract_cog_radar}. For the multiple constraint case, the solution to $\argmax~G_\time(\boldsymbol{\lambda})$ is the vector of Lagrange multipliers for the constraints at $\response_\time$~\eqref{eqn:abstract_cog_radar}, the optimal response at time $\time$. Denoting the IRL feasibility test wrt inequalities~\eqref{eqn:abstract_IRL_utility_vec_g} and \eqref{eqn:not_all_zeros_vec_g} as $\boldsymbol{\AFT}(\param,\dataset_\utility)$, the margin $\margin_\utility(\dataset_\utility)$ for any utility $\utility$ when $\nonlinbrad$ is vector-valued can be compactly defined as:
\begin{equation}\label{eqn:margin_rp_vec_g_abstract}
    \margin_\utility(\dataset_\nonlinb) = \min_{\eps\geq 0}~\eps,~\boldsymbol{\AFT}(\param,\dataset_g) + \eps\mathbf{1}\geq \mathbf{0}.
\end{equation}

Having generalized the margin definition of \eqref{eqn:margin_rp} in Definition~\ref{def:margin_utility} to the multiple constraint case, we now state our cognition masking result, Theorem~\ref{thrm:irp_vec_g} for vector-valued $\nonlinbrad$. The cognition masking rationale for vector-valued $\nonlinbrad$ remains the same as that in Theorem~\ref{thrm:irp}: add engineered noise to the radar's optimal responses, and ensure the radar's utility $\utilityrad$ lies sufficiently close to the edge of the feasibility polytope of viable utilities computed via IRL. 

\begin{theorem}[Masking Utility from Adversarial IRL for Multiple Resource Constraints.]
Consider the cognitive radar~\eqref{eqn:abstract_cog_radar} from Definition~\ref{def:cognitive-radar} with multiple resource constraints (assumption~\ref{asmp:utility_vec_g} holds).
Let $\{\response_\time^\ast\}_{\time=1}^\horizon$ denote the naive response sequence~\eqref{eqn:abstract_cog_radar} that maximizes the cognitive radar's utility.
The response sequence $\{\pertresponse_{1:\horizon}^\ast\}$ defined below masks the radar's utility $\utilityrad$ from the adversary by ensuring $\utilityrad$ satisfies the IRL inequalities~\eqref{eqn:abstract_IRL_utility_vec_g},~\eqref{eqn:not_all_zeros_vec_g} with a sufficiently low margin~\eqref{eqn:margin_rp_vec_g}:\vspace{-0.2cm}
\begin{align}
 \hspace{-0.5cm}\{\pertresponse_{1:\horizon}^\ast\}& = \underset{\{\response_\time\geq \mathbf{0},~ \nonlinbrad(\probe_\time,\response_\time)\leq \mathbf{0} \}}{\argmin} \sum_{\time=1}^\horizon \utilityrad(\idresponse_\time) - \utilityrad(\response_\time), \label{eqn:irp_vec_g}\\
&\margin_\utilityrad(\dataset_\nonlinb) \leq (1-\eta)~\margin_\utilityrad(\dataset_\nonlinb^\ast),\label{eqn:constraint_lowmargin_vec_g}
\end{align}
where  dataset $\dataset_\nonlinb^\ast=\{\nonlinbrad(\cdot),\idresponse_\time\}_{\time=1}^\horizon$ is the adversary's dataset when the radar transmits naive responses $\{\idresponse_\time\}_{\time=1}^\horizon$, and $\dataset_\nonlinb$ is defined in \eqref{eqn:dataset_IRL_utility}.
\label{thrm:irp_vec_g}
\end{theorem}
The proof of Theorem~\ref{thrm:irp_vec_g} is straightforward and omitted due to brevity. The only distinguishing factor between Theorems~\ref{thrm:irp_vec_g} and \ref{thrm:irp} is the generalized definition of the margin $\margin_\utilityrad$~\eqref{eqn:margin_rp_vec_g} for vector-valued constraints.

\subsection{Cognition Masking Performance under Misspecified Radar Response Measurements}\label{appdx:misspec}
In this section, we investigate how the performance of the radar's cognition masking algorithm, Theorem~\ref{thrm:irp}, changes when the adversary has misspecified measurements of the radar's responses. By misspecified responses, we mean the true radar response $\response_\time$ is corrupted by an additive deterministic perturbation $\misspec_\time$, $\time=1,2,\ldots,\horizon$. Misspecified response measurements formalized in assumption \ref{asmp:misspec} below are different from noisy measurements since the perturbation $\misspec_\time$ is a deterministic vector and not a random variable like in the noisy case considered in Sec.\,\ref{sec:stoch_irl}. 
\begin{assumption}[Misspecified Radar Response Measurements]
    Suppose the adversary has misspecified measurements $\bar{\response_\time}=\response_\time + \misspec_\time,~\misspec_\time\in\reals^\probedim$ of the radar's response $\response_\time$. Assume the misspecifications $\zeta_\time$ have a bounded $\mathcal{L}_2$-norm:
    \begin{equation}\label{eqn:misspec-bound-l2}
    \|\misspec_\time\|_2 \leq \mathbf{\misspec}
    \end{equation}
    The adversary's misspecified datasets $\bar{\dataset}_\nonlinb$ and $\bar{\dataset}_\utility$ are defined as:
    \begin{equation}
        \label{eqn:misspec-dataset}
        \begin{split}
            \bar{\dataset}_\nonlinb & \equiv \{\nonlinbrad(\probe_\time,\cdot),\response_\time + \misspec_\time\}_{\time=1}^\horizon,~\misspec_\time\in\reals^\probedim\\
            \bar{\dataset}_\utility & \equiv \{\utilityrad(\probe_\time,\cdot),\response_\time + \misspec_\time\}_{\time=1}^\horizon,~\misspec_\time\in\reals^\probedim
        \end{split}
    \end{equation}
where $\response_\time$ is the radar's response at time $\time$, $\utilityrad$ and $\nonlinbrad$ are the utility and constraint, respectively, of the cognitive radar~\eqref{eqn:abstract_cog_radar}.
    \label{asmp:misspec}
\end{assumption}

Recall from Theorem~\ref{thrm:irp} that the positive scalar $\eta$ parametrizes the extent of cognition masking by the cognitive radar. Our key objective is to derive a lower bound on the effective extent of cognition masking $\eta_{\operatorname{eff}}$ defined as:
\begin{equation}
\begin{split}
    &\hspace{-0.4cm}\eta_{\operatorname{eff}} = \margin_\utilityrad(\widetilde{\bar{\dataset}}_\nonlinb)/\margin_\utilityrad(\bar{\dataset}_\nonlinb)~(\text{when assumption~\ref{asmp:utility} holds}),\\
    &\hspace{-0.4cm}\eta_{\operatorname{eff}} = \margin_\nonlinbrad(\widetilde{\bar{\dataset}}_\utility)/\margin_\nonlinbrad(\bar{\dataset}_\utility)~(\text{when assumption~\ref{asmp:constraint} holds}),
\end{split}
\label{eqn:eta-effective}
\end{equation}
where $\bar{\dataset}_\nonlinb,\bar{\dataset}_\utility$~\eqref{eqn:misspec-dataset} are the  misspecified datasets of the adversary if the radar transmits naive responses~\eqref{eqn:abstract_cog_radar}, and $\widetilde{\bar{\dataset}}_\nonlinb,\bar{\dataset}_\utility$ are the misspecified datasets when the radar transmits cognition masking responses computed via \eqref{eqn:irp} and \eqref{eqn:irp_constraint}, respectively. The variable $\eta_{\operatorname{eff}}$~\eqref{eqn:eta-effective} is the ratio between the margins of the radar's strategy (utility $\utilityrad$ or constraint $\nonlinbrad$)~\eqref{eqn:margin_rp},~\eqref{eqn:margin_rp_constraint} with and without the radar's cognition masking scheme (Theorem~\ref{thrm:irp}) when the adversary has misspecified response measurements. It is easy to see that $\eta_{\operatorname{eff}}=\eta$ if the misspecification error $\misspec_\time$~\eqref{eqn:misspec-dataset} is $0$ for all $\time$. For non-zero $\misspec_\time$, Theorem~\ref{thrm:misspec} below yields a lower bound for $\eta_{\operatorname{eff}}$ and uses the following variables (given assumption~\ref{asmp:misspec} holds):
\begin{equation}
\begin{split}
    d_{1,\utilityrad} & = \min_{\time}  \nabla \utilityrad(\response_\time)'\misspec_\time~,~d_{2,\utilityrad} = \max_{\time}  \nabla \utilityrad(\response_\time)'\misspec_\time,\\
    d_{1,\nonlinbrad} & = \min_{\time}  \nabla \nonlinbrad(\response_\time)'\misspec_\time~,~d_{2,\nonlinbrad} = \max_{\time}  \nabla \nonlinbrad(\response_\time)'\misspec_\time,
\end{split}
\label{eqn:spread-misspec}
\end{equation}
The variables defined in \eqref{eqn:spread-misspec} measure the deviation in the radar's utility and constraint values evaluated at the misspecified radar responses measured by the adversary, compared to the utility and constraint evaluations at the true radar responses. We are now ready to state Theorem~\ref{thrm:misspec}.
\begin{theorem}[Performance of Cognition Masking (Theorem~\ref{thrm:irp}) for Misspecified Responses] \label{thrm:misspec}
Consider the cognition masking scheme of Theorem~\ref{thrm:irp}. Assume the adversary has misspecified radar response measurements (assumption \ref{asmp:misspec} holds).  
Then:\\
(i) Suppose assumption \ref{asmp:utility} holds, i.e.\,, the adversary knows the radar's constraint $\nonlinbrad$. Then, the effective extent of cognition masking $\eta_{\operatorname{eff}}$ is bounded from below as:
\begin{equation}\label{eqn:misspec-bound-u}
    \eta_{\operatorname{eff}} \geq \eta - \left(\frac{(1-\eta)~(d_{2,\utilityrad} - d_{1,\utilityrad})}{\margin_\utilityrad(\dataset_\nonlinb) - d_{2,\utilityrad}}\right)
\end{equation}
(ii) Suppose assumption \ref{asmp:constraint} holds, i.e.\,, the adversary knows the radar's constraint $\utilityrad$. Then, the effective extent of cognition masking $\eta_{\operatorname{eff}}$ is bounded from below as:
\begin{equation}\label{eqn:misspec-bound-g}
    \eta_{\operatorname{eff}} \geq \eta - \left(\frac{(1-\eta)~(d_{2,\nonlinbrad} - d_{1,\nonlinbrad})}{\margin_\nonlinbrad(\dataset_\utility) - d_{2,\nonlinbrad}}\right)
\end{equation}    
The variables $d_{1,\utilityrad},d_{2,\utilityrad},d_{1,\nonlinbrad},d_{2,\nonlinbrad}$ measure the distortion in the adversary's dataset due to misspecified measurements and defined in \eqref{eqn:spread-misspec}.
\end{theorem}
Theorem~\ref{thrm:misspec} computes a lower bound on the effectiveness of the cognitive masking scheme of Theorem~\ref{thrm:irp} when the adversary has misspecified measurements of the radar's response. The proof for Theorem~\ref{thrm:misspec} is omitted for brevity. Observe that the bounds in \eqref{eqn:misspec-bound-u},~\eqref{eqn:misspec-bound-g} are inversely proportional to the quantities $(d_{2,\utilityrad}-d_{1,\utilityrad})$ and $(d_{2,\nonlinbrad}-d_{1,\nonlinbrad})$. These quantities can be interpreted as the `spread' in the utility and constraint evaluations at the radar's true responses due to the misspecification errors $\misspec_\time$~\eqref{eqn:misspec-dataset} and, in turn, are proportional to $\mathbf{\misspec}$~\eqref{eqn:misspec-bound-l2}, the maximum $\mathcal{L}_2$ norm of $\{\misspec_{\time}\}_{\time=1}^\horizon$. Hence, we can conclude the lower bound for the effectiveness of the radar's cognition masking scheme worsens with the magnitude of the misspecification errors in the adversary's measurements.

\subsection{Cognition Masking for Arbitrary IRL Algorithm}\label{appdx:arb-irl}
Our cognition masking results of Theorems~\ref{thrm:irp} and \ref{thrm:noise_irp} assume the adversary performs optimal IRL via Afriat's theorem (Theorems~\ref{thrm:rp} and \ref{thrm:rp_constraint}) to reconstruct the radar's strategy. However, our cognition masking results can be straightforwardly extended to any IRL algorithm. Any IRL algorithm can be expressed WLOG as a set-valued estimation algorithm that generates a set of feasible strategies given a finite dataset of adversary probes $\{\probe_\time\}_{\time=1}^\horizon$ and radar responses $\{\response_\time\}_{\time=1}^\horizon$:
\begin{equation}\label{eqn:arb-irl-notation}
    \hspace{-0.3cm}\IRLalg(\param,\{\probe_\time,\response_\time\}_{\time=1}^\horizon)\leq \mathbf{0}, \text{where}\IRLalg: \Theta \times \reals_{+}^{2\probedim\horizon} \rightarrow \reals^\ineqdim
\end{equation}
In \eqref{eqn:arb-irl-notation}, $\param\in\Theta$ parametrizes the reconstructed utility, $\{\probe_\time,\response_\time\}_{\time=1}^\horizon$ is the adversary's dataset and $L$ is the number of IRL feasibility inequalities. In Afriat's theorem, for example, $\Theta=\reals_+^{2\horizon}$ and $\ineqdim=\horizon^2-\horizon$. Algorithm~\ref{alg:arb-irl} below outlines the steps for mitigating an arbitrary IRL algorithm $\IRLalg(\cdot,\{\probe_\time,\response_\time\}_{\time=1}^\horizon)$. Recall Theorem~\ref{thrm:irp} minimizes the feasibility margin of the radar's strategy wrt the Afriat inequalities~\eqref{eqn:abstract_IRL_utility},~\eqref{eqn:abstract_IRL_constraint} by deliberately perturbing the radar's responses. In complete analogy, a radar can hide its strategy from any set-valued IRL estimation scheme by minimizing the feasibility margin defined below in \eqref{eqn:margin_arb_IRL} wrt the IRL feasibility inequalities $\IRL(\cdot,\{\probe_\time,\response_\time\}_{\time=1}^\horizon)$ by purposefully injecting noise in the radar's responses. Due to the non-linear margin constraint in \eqref{eqn:implementation-iirl-arb}, the optimization problem can be solved using a general purpose non-linear programming solver, for example, \verb|fmincon| in MATLAB, to obtain a local minimum.

\begin{algorithm}\caption{Masking Radar Utility from Arbitrary IRL algorithm $\IRLalg(\cdot,\{\probe_\time,\response_\time\}_{\time=1}^\horizon)\leq \mathbf{0}$ }\label{alg:arb-irl}
    Step 1. Compute radar's naive response sequence $\idresponse_{1:\horizon}$ by solving the convex optimization problem~\eqref{eqn:abstract_cog_radar}:
\begin{equation*}
    \idresponse_\time = \argmin \utilityrad(\response),~\nonlinbrad(\probe_\time,\response)\leq 0,~\response\geq\mathbf{0}~\forall\time\in\{1,2,\ldots,\horizon\},
\end{equation*}
where $\utilityrad$ is concave monotone in $\response$ and $\nonlinbrad(\probe_\time,\response)$ is convex monotone in $\response$.\\
Step 2. Choose $\eta\in[0,1]$ (extent of cognition masking from IRL feasibility test).\\
Step 3. Compute the margin of the naive responses wrt the IRL algorithm:
\begin{align}
   \margin_{\utility}(\{\probe_\time,\response_\time^\ast\}_{\time=1}^\horizon;&\IRLalg) = \min_{\eps\geq 0}~\eps,\nonumber\\
   &\IRLalg(\utilityrad,\{\probe_\time,\response_\time^\ast\}_{\time=1}^\horizon) +\eps\mathbf{1}\geq\mathbf{0},\label{eqn:margin_arb_IRL}
\end{align}
where $\mathbf{1}$ is a vector of all ones and $\utilityrad$ is the radar's utility.\\
Step 3. Compute upper bound $\margin_{\operatorname{thresh}}$ on desired margin~\eqref{eqn:margin_rp} after cognition masking:\\
$\margin_{\operatorname{thresh}}=(1-\eta)~\margin_\utilityrad(\{\probe_\time,\idresponse_\time\}_{\time=1}^\horizon;\IRLalg)$.\\
Step 4. Compute the cognition-making response sequence $\{\pertresponse_{1:\horizon}^\ast\}$ by solving the following optimization problem:
\begin{equation}\begin{split}
&\quad\min~\sum_{\time=1}^\horizon \utilityrad(\idresponse_\time) - \utilityrad(\response_\time),\\
&\response_\time\geq \mathbf{0},~\probe_\time'\response_\time\leq 1~\forall\time\in\{1,2,\ldots,\horizon\},\\
&\margin_{\utilityrad}(\{\probe_\time,\response_\time\}_{\time=1}^\horizon;\IRLalg)  \leq\margin_{\operatorname{thresh}}.
\end{split}
\label{eqn:implementation-iirl-arb}
\end{equation}
\end{algorithm}

\bibliographystyle{unsrt_abbrv_custom} 
\bibliography{refs}



\end{document}

%% file: macrodefs.tex
\newcommand{\thresh}{\gamma}
\newcommand{\prob}{\mathbb{P}}
\newcommand{\empprob}{\widehat{\prob}}

\newcommand{\probe}{\alpha}
\newcommand{\response}{\beta}
\newcommand{\utility}{u}

\newcommand{\reals}{\mathbb{R}}
\newcommand{\dataset}{\mathcal{D}}

\newcommand{\pdf}{p}
\newcommand{\belief}{\pi}
\newcommand{\state}{x}
\newcommand{\obs}{y}

\newcommand{\statenoisecov}{Q}

\renewcommand{\time}{k}
\newcommand{\filter}{\mathcal{T}}
\newcommand{\horizon}{K}

\newcommand{\probedim}{m}
\newcommand{\nresponse}{\hat{\response}}
\newcommand{\pertresponse}{\tilde{\response}}

\newcommand{\idresponse}{\response^\ast}
\newcommand{\ndataset}{\widehat{\dataset}}  
\newcommand{\vecresponse}{\boldsymbol{\response}}

\newcommand{\siglevel}{\gamma}
\newcommand{\numtargets}{\probedim}

\newcommand{\iter}{i}

\newcommand{\normal}{\mathcal{N}}
\newcommand{\trace}{\operatorname{Tr}}
\newcommand{\snoisecov}{Q}
\newcommand{\onoisecov}{R}
\newcommand{\statem}{A}
\newcommand{\obsm}{C}
\newcommand{\snoise}{w}
\newcommand{\onoise}{v}
\newcommand{\argmin}{\operatorname{argmin}}
\newcommand{\argmax}{\operatorname{argmax}}
\newcommand{\nonlinb}{g}

\newcommand{\margin}{\operatorname{\mathcal{M}}}

\newcommand{\eps}{\epsilon}

\newcommand{\IRL}{\operatorname{IRL}}
\newcommand{\AFT}{\operatorname{\mathcal{A}}}
\newcommand{\param}{\theta}
\newcommand{\paramcons}{\theta}

\newcommand{\IIRLU}{\operatorname{MASK-U}}

\newcommand{\utilityrad}{\mathbf{\utility}}
\newcommand{\nonlinbrad}{\mathbf{\nonlinb}}
\newcommand{\gradstep}{\delta}
\newcommand{\misspec}{\zeta}
\newcommand{\resnoise}{\omega}
\newcommand{\IRLalg}{\operatorname{IRL}}
\newcommand{\gdim}{I}
\newcommand{\ineqdim}{L}
\newcommand{\paramg}{\kappa}